\documentclass[final,5p,times,twocolumn]{elsarticle_nofoot}
\usepackage{subfigure}
\usepackage[utf8]{inputenc}
\usepackage{hyperref}
\usepackage{xspace}
\hypersetup{colorlinks=true,%
           linkcolor=black,%
           anchorcolor=black,%
           citecolor=black,%
           filecolor=black,%
           menucolor=black,%
           urlcolor=black}
\biboptions{comma,square,sort&compress}
\newcommand{\neqcm}{\ensuremath{\mathrm{n}_{\mathrm{eq}}/\mathrm{cm}^2}\xspace}
\newcommand{\mum}{\ensuremath{\mu}m\xspace}
\newcommand{\mumq}{\ensuremath{{{\mu}\mathrm{m}}^2}\xspace}
\newcommand{\degC}{\ensuremath{^{\circ}\mathrm{C}}\xspace}
\newcommand\fig{Figure}

\newcommand\tab{Table} 
\newcommand{\Am}{\ensuremath{^{241}}Am\xspace}

\newcommand{\Sr}{\ensuremath{^{90}}Sr\xspace}
\newcommand{\dactive}{\ensuremath{d_{\mathrm{active}}}\xspace}
\newcommand{\dinactive}{\ensuremath{d_{\mathrm{inactive}}}\xspace}
\newcommand{\dthin}{\ensuremath{\dactive=75}\,\mum}

\journal{Nuclear Instruments and Methods A}
\begin{document}
\begin{frontmatter}
%
\title{Production and Characterisation of SLID Interconnected n-in-p Pixel
  Modules with 75 Micrometer Thin Silicon Sensors}
\author[label2]{L.~Andricek}
\author[label1]{M.~Beimforde}
\author[label1]{A.~Macchiolo}
\author[label1]{H-G.~Moser}
\author[label1]{R.~Nisius\corref{cor1}}
\ead{Richard.Nisius@mpp.mpg.de}
\cortext[cor1]{Corresponding author}
\author[label2]{R.H.~Richter}
\author[label1]{S.~Terzo}
\author[label1]{P.~Weigell}
\address[label1]{Max-Planck-Institut f\"ur Physik (Werner-Heisenberg-Institut),
  F\"ohringer Ring 6, D-80805 M\"unchen, Germany}
\address[label2]{Halbleiterlabor der Max-Planck-Gesellschaft, Otto Hahn Ring 6,
  D-81739 M\"unchen, Germany}
%
\begin{abstract}
 The performance of pixel modules built from 75 micrometer thin silicon sensors
 and ATLAS read-out chips employing the Solid Liquid InterDiffusion (SLID)
 interconnection technology is presented.
 This technology, developed by the Fraunhofer EMFT, is a possible alternative to
 the standard bump-bonding. It allows for stacking of different interconnected
 chip and sensor layers without destroying the already formed bonds. In
 combination with Inter-Chip-Vias (ICVs) this paves the way for vertical
 integration. Both technologies are combined in a pixel module concept which is
 the basis for the modules discussed in this paper.

 Mechanical and electrical parameters of pixel modules employing both SLID
 interconnections and sensors of 75 micrometer thickness are covered. The
 mechanical features discussed include the interconnection efficiency, alignment
 precision and mechanical strength. The electrical properties comprise the
 leakage currents, tuning characteristics, charge collection, cluster sizes and
 hit efficiencies. Targeting at a usage at the high luminosity upgrade of the
 LHC accelerator called HL-LHC, the results were obtained before and after
 irradiation up to fluences of $10^{16}$\,\neqcm.
\end{abstract}
%
%
\end{frontmatter}
%
%
\section{Future Pixel Modules and 3D-Integration Technology}
 The ATLAS pixel detector~\cite{pixelelectronics} is made of three barrel layers
 with an innermost radius of 50.5\,mm and three end-cap discs on each side of
 the detector.
 The pixel modules used consist of 16 FE-I3 read-out chips~\cite{Peric2006178}
 which are interconnected via the solder bump bonding
 technique~\cite{Fritzsch2011189} to a 250\,\mum\ thick n-in-n planar silicon
 sensor.
 The size of individual pixel cells is 50\,\mum\ $\times$\,400\,\mum. Sensors
 and read-out chips are specified for a maximum fluence\footnote{The fluences
   for proton and neutron irradiation are rescaled to the damage expected for
   1\,MeV neutrons, indicated by \neqcm.} of $10^{15}$\,\neqcm\ and a dose of
 500\,kGy.

 A large upgrade to the LHC accelerator chain - called HL-LHC - is currently
 planned to start taking data in 2024. 
 The peak luminosity will eventually be increased up to
 5$\cdot10^{34}$\,cm$^{-2}$s$^{-1}$~\cite{Lumi}. To maintain the detector
 performance, several upgrades of the ATLAS pixel detector are planned.
 The first of these upgrades, the so called Insertable B-Layer
 (IBL)~\cite{IBL-TDR}, is a new fourth pixel layer, which is planned to be
 mounted on a new smaller beam pipe at a radius of 32\,mm, and to be operational
 by the end of 2014. Due to the smaller radius, the modules cannot overlap along
 the beam direction as they do for the present ATLAS pixel detector. Thus, the
 active fraction of the new pixel modules was
 increased~\cite{BenoitPhD,WittigPhD,IBL_Proto}. Additionally, the harsher
 radiation environment and the higher occupancy demanded for a new read-out
 chip, the FE-I4~\cite{GarciaSciveres2010}, specified up to a received fluence
 of $5\cdot10^{15}$\,\neqcm\ and with a reduced pixel size of
 50\,\mum$\,\times\,$250\,\mum. Furthermore, the number of pixel cells increased
 from 2880 to 26880 per chip. While it is expected that the upgraded pixel
 detector retains sufficient tracking capabilities until around 2024, a full
 replacement of the tracking detector is required afterwards.

 The current baseline detector upgrade layout~\cite{LOI_II} consists of four
 pixel layers at a minimal radius of about 39\,mm, supplemented by six pixel
 discs at each of the forward regions, extending to a pseudo-rapidity of about
 $\pm 2.8$.
 Given the corresponding extreme radiation levels of up to
 $2\cdot10^{16}$\,\neqcm\ in the innermost layer a new generation of read-out
 chips will be needed for the inner layers, featuring even smaller pixels to
 cope with the otherwise largely increased pixel occupancy.
%
%
\subsection{Module Concept}
%
\begin{figure*}[tbh]
\centering
\subfigure[]{
\includegraphics[width=0.47\textwidth]{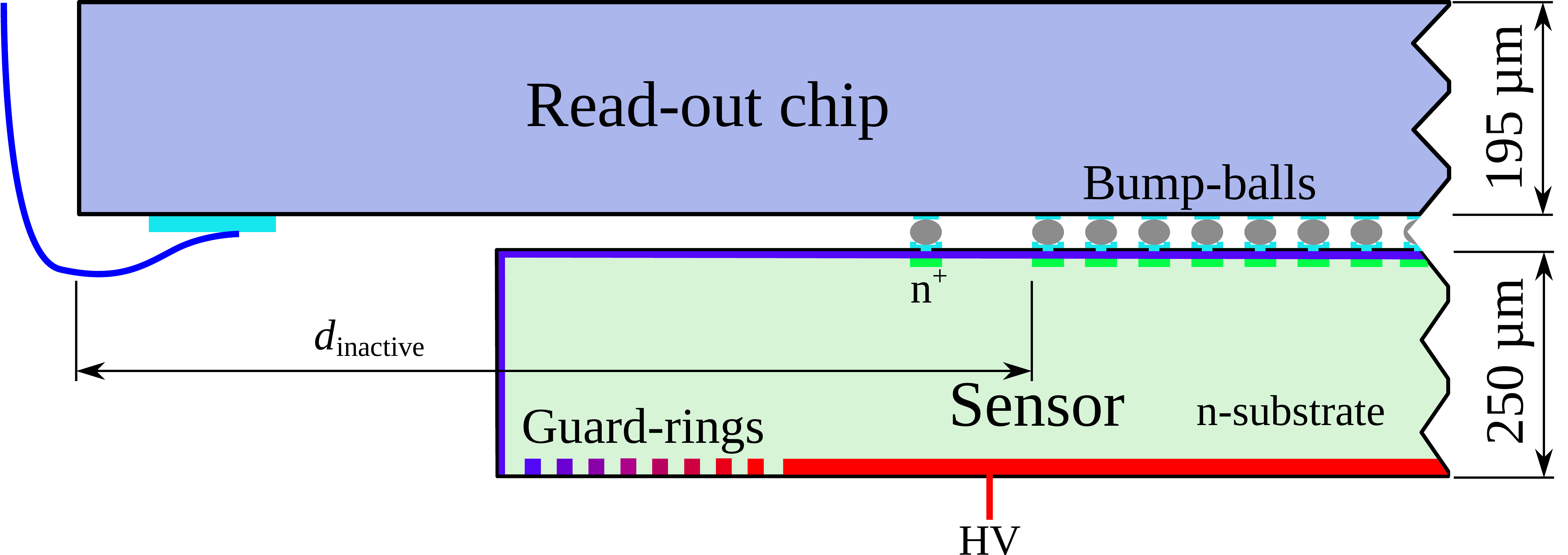}
\label{fig:CurrentDesign}
}
\subfigure[]{
\includegraphics[width=0.47\textwidth]{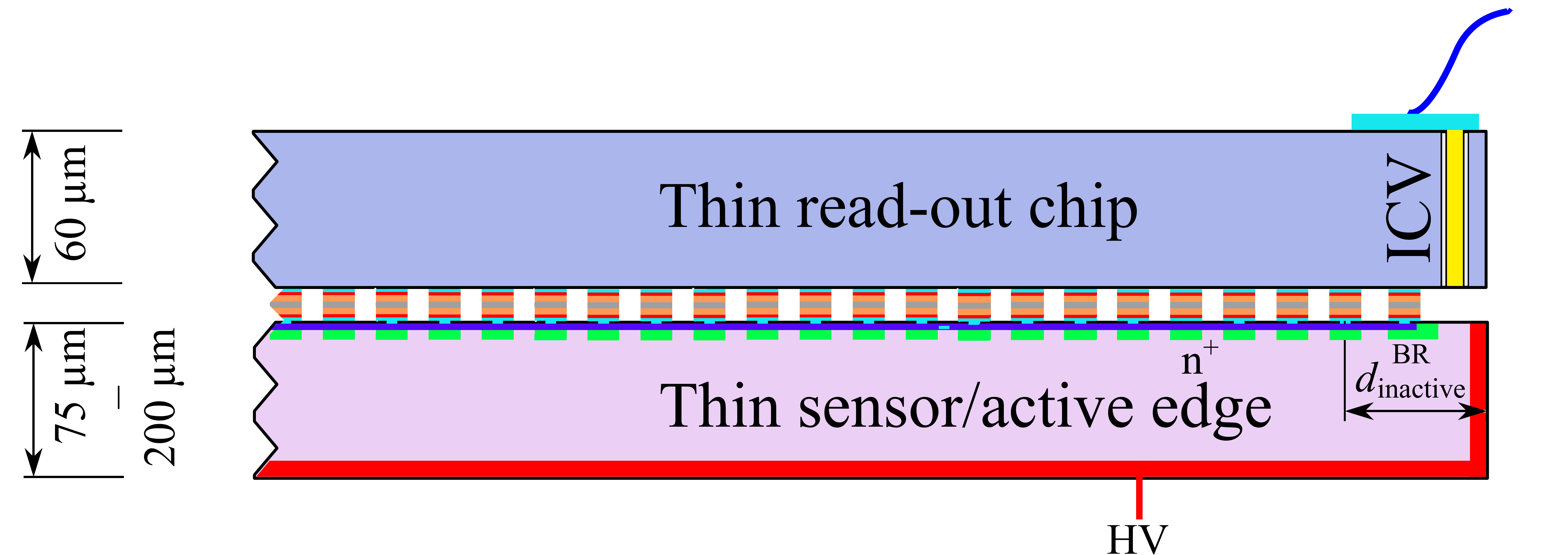}
\label{fig:NewConcept}
}
\caption[Comparison sketch of the current and proposed new
  design.]{\subref{fig:CurrentDesign} Schematics of the current pixel detector
  module within the ATLAS experiment and \subref{fig:NewConcept} the
  investigated design. The read-out chip is shown in blue, the sensor in light
  green (violet) for the current (new) concept. The pixel implants as well as
  the bias ring (BR) n$^+$-implants are indicated in green, where the implant
  closest to the edge is the bias ring implant. The back-side implantation, and
  for the new concept also the edge implantation, is indicated in red. Wire bond
  pads are drawn in light blue, ICVs in yellow.
 The symbol \dinactive\ denotes the distance from the last read-out pixel to the
 edge of the module. (For interpretation of the references to colour in this
 figure caption, the reader is referred to the web version of this paper.)
  \label{fig:ConceptComparison}}
\end{figure*}
%
 To answer the challenges of this upgrade, a module concept for the pixel layers
 is investigated, which employs several novel technologies in the field of pixel
 detectors: n-in-p pixel sensors are thinned using a process~\cite{Lacithin}
 developed at the Max-Planck-Gesellschaft Halbleiterlabor~(MPG-HLL), and
 connected via the Fraunhofer EMFT~\cite{EMFT} Solid Liquid Inter-Diffusion
 (SLID) technology to the read-out electronics, where the signals are routed via
 Inter-Chip-Vias (ICVs). Additionally, the active fraction is maximised by an
 optimised guard ring design in combination with or without implanted sensor
 sides~\cite{TerzoIWORID}. In \fig~\ref{fig:ConceptComparison} the schematics of
 \subref{fig:CurrentDesign} the present and \subref{fig:NewConcept} the
 investigated concept are shown.

 Advantages of this approach are: the {n-in-p} technology allows for single
 sided processing of wafers resulting in a lower cost, which is of special
 importance for the large areas foreseen in future pixel detector upgrades.
 In addition, the radiation hardness is comparable to the presently used
 {n-in-n} technology~\cite{NinPpaper,WeigellPhD}. Thinner sensors not only
 reduce the material budget and therefore multiple scattering but also, at the
 same applied bias voltage, they exhibit higher electric fields than thicker
 devices. This leads to a high charge collection efficiency (CCE) after high
 radiation doses at moderate bias
 voltages~\cite{Casse2010401,Mandic2010474,WeigellPhD}. While on the sensor side
 the inactive area is removed by activated edges, in a 3D compliant design of
 the pixel electronics, ICVs could eventually avoid the need for the cantilever
 area where presently the wire bonding pads are located.
 Combining these ICVs with SLID interconnections enables fully 3D-integrated
 modules. Furthermore, SLID interconnections could allow for a pitch reduction
 with respect to the 50\,\mum\ pitch limit given by the solder bump bonding.

 In this paper, the results on the SLID interconnection and thin sensor aspects
 of this module concept are presented and discussed.
 Some preliminary results were already given
 in~\cite{AnnaJapan,AnnaTippSLID,WeigellPsd,AnnaGrindel,BeimfordeTWEPP} as well
 as in the PhD-theses~\cite{BeimfordePhD,WeigellPhD}.
 All SLID modules presented here use the FE-I2~\cite{FE-I2} ATLAS readout chip
 that has the same footprint as the FE-I3 chip. The two chips only differ in
 minor details, e.\,g. they need slightly different chip analogue and digital
 voltages documented in~\cite{stockmanns}. These differences are not relevant
 for the work presented in this paper. Consequently, in the following no
 distinction is made and both chips are referred to as FE-I3 chips.
 Further results on the other technologies used in the module concepts, as
 n-in-p sensors, active edge pixel devices and ICVs, can be found
 in~\cite{Weigell2011,TerzoIWORID,WeigellPhD,BeimfordePhD}.

 In the presentation of the results first a short introduction to SLID will be
 given, followed by the technical and mechanical results. Finally, the
 performance for prototype pixel modules employing SLID and 75\,\mum\ thick
 sensors will be discussed.
%
%
\subsection{Solid-Liquid InterDiffusion}
\label{sect:SLID}
%
\begin{figure*}[tbh]
\centering
\subfigure[\hspace{1.25cm}]{
\includegraphics[width=0.23\textwidth]{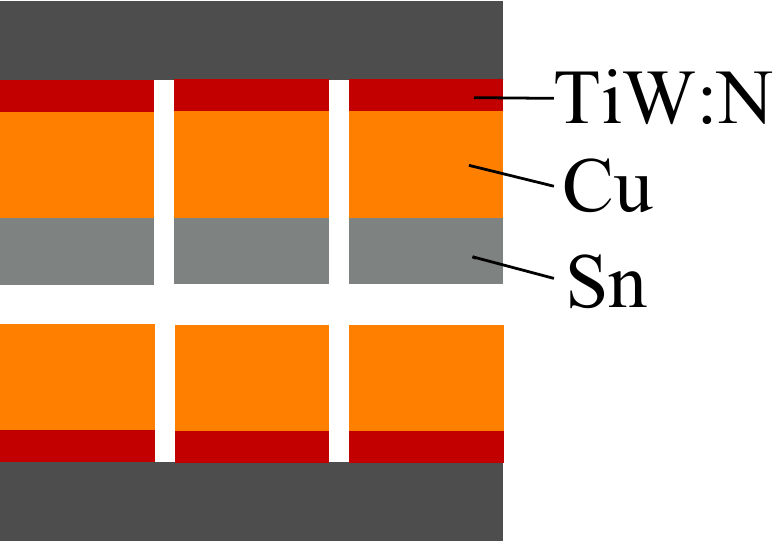}
\label{fig:SLIDProcess_a}
}
\subfigure[\hspace{1.25cm}]{
\includegraphics[width=0.23\textwidth]{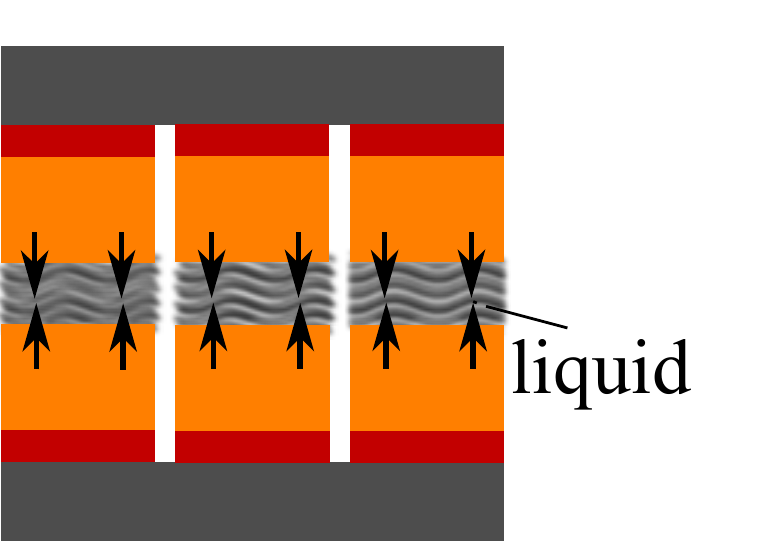}
\label{fig:SLIDProcess_b}
}
\subfigure[\hspace{1.25cm}]{
\includegraphics[width=0.23\textwidth]{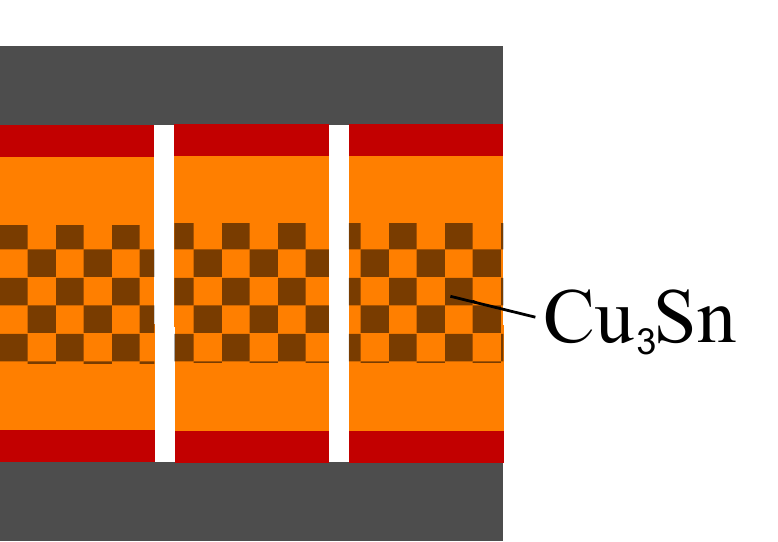}
\label{fig:SLIDProcess_c}
}
\subfigure[]{
\includegraphics[width=0.20\textwidth]{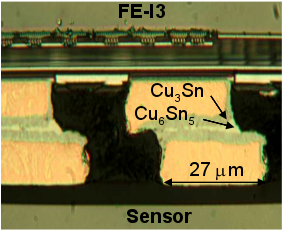}
\label{fig:SLIDcut}
}
\caption[Process flow of the SLID interconnection and cross section of SLID pads
  in a pixel module.]{(a-c) Process flow of the SLID interconnection, adapted
  from~\cite{Klumpp3DBuch}. For an explanation of the individual steps please
  refer to the text. \subref{fig:SLIDcut} Cross section of SLID pads in a pixel
  module built from an FE-I3 read-out chip and a dummy sensor (aluminium and
  silicon oxide only). Photograph taken by~\cite{EMFT}.
  \label{fig:SLID_Process}}
\end{figure*}
%
 SLID is a class of interconnection techniques, where the formation of the
 interconnection takes place at temperatures significantly lower than those the
 connections can tolerate afterwards without dissolving. The concept was
 introduced in the 1960s~\cite{Bernstein1966,Bernstein1965} and is based on
 binary, ternary, or even higher-order metal systems, where one low-temperature
 melting metal is coated on a high-temperature melting core. By bringing the
 temperature of the metal system above the melting point of the low-temperature
 melting metal and applying high pressure, this metal dissolves and diffuses
 into the high-temperature melting metal. Inter-metallic compounds with melting
 points above the heating temperature are formed by the two metals and the
 liquid phase solidifies.
 While many different metal systems are known to form {SLID} bonds, certain
 constraints apply when using this technique in real
 applications~\cite{Klumpp3DBuch}. For example, the melting point of the
 low-temperature melting metal should be below 400\,\degC, which is the maximum
 temperature most Application Specific Integrated Circuits (ASICs) can
 withstand. In the presented module concept, the {SLID} process developed by the
 {EMFT} is used. The process steps are shown in \fig~\ref{fig:SLID_Process}.  In
 this approach Sn ($T_{\mathrm{melt}}=231.9$\,\degC) is used as the
 low-temperature melting component and Cu ($T_{\mathrm{melt}}=1083.0$\,\degC) as
 the high-temperature melting component.
 Out of these, Cu$_3$Sn ($T_{\mathrm{melt}}=676$\,\degC) and Cu$_6$Sn$_5$
 ($T_{\mathrm{melt}}=415$\,\degC) are formed. The high melting point of the
 interconnecting alloy opens the possibility of subsequent stacking of
 additional {SLID}-interconnected layers, but at the same time inhibits
 reworking of badly connected devices.

\begin{figure}[tbh]
  \begin{center}
    \includegraphics[width=0.5\textwidth]{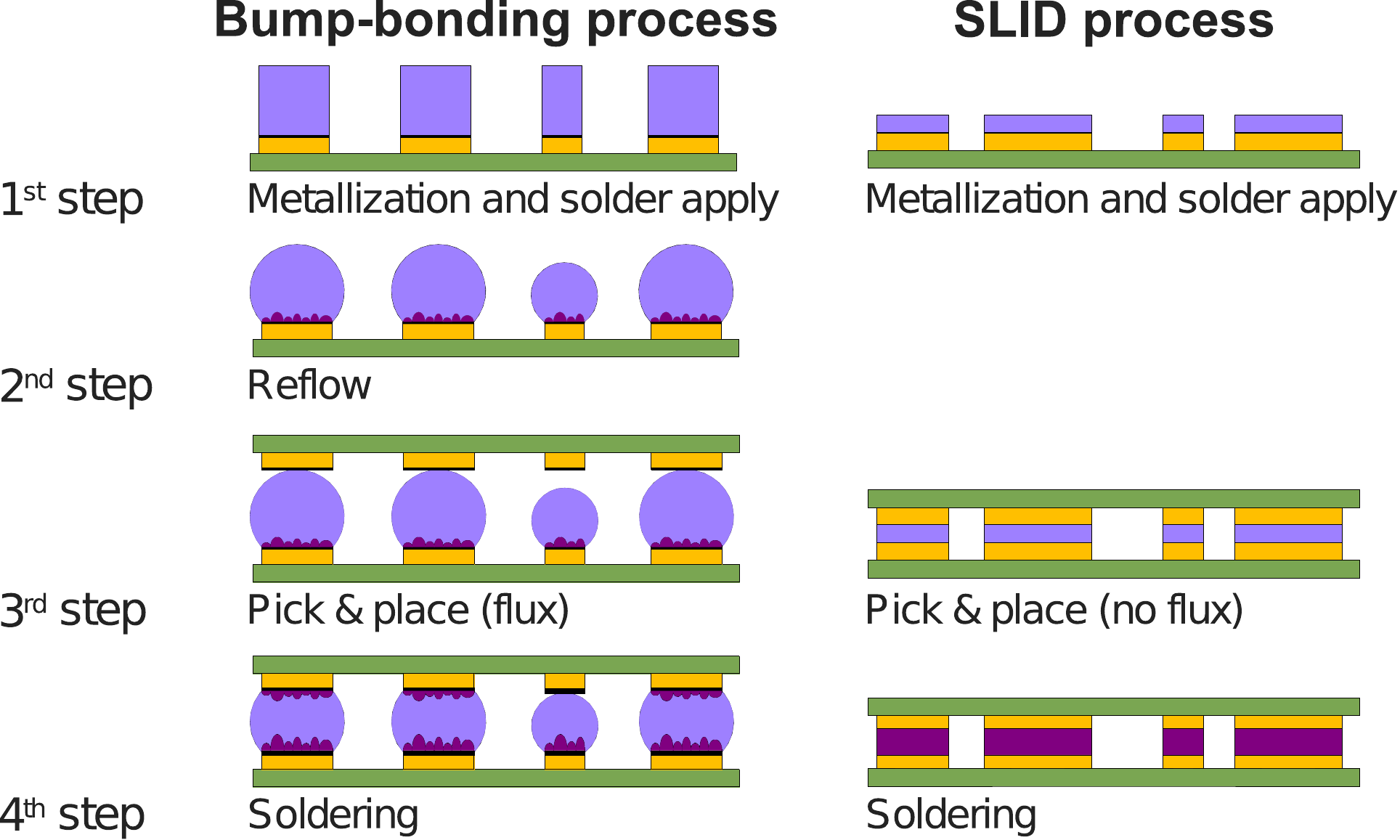}
    \caption[Step-by-step comparison of the bump-bonding and SLID
      interconnection technologies.]{Step-by-step comparison of the bump-bonding
      and SLID interconnection technologies~\cite{EMFT}.}
    \label{fig:ProcessFlowCompare}
  \end{center}
\end{figure}

 A comparison of the process flows of the conventional bump bonding techniques
 and {SLID}, shown in \fig~\ref{fig:ProcessFlowCompare}, reveals further
 advantages and challenges. While the first step is a patterned electroplating
 step needed for the deposition of the Cu and Sn, which is similar for both
 technologies, the so-called reflow step is not needed to form {SLID}
 interconnections. In the reflow step, the alloy or metal, e.\,g.\ PbSn or In
 that are used in the bump bonding process is melted; the surface tension leads
 to solder ball formation. Since the diameter of the balls is determined by the
 initial pad size, all bump-bond connections have to be of equal size to form
 good connections (compare 3$^{\mathrm{rd}}$ connection to the neighbouring ones
 in \fig~\ref{fig:ProcessFlowCompare}). In contrast, a {SLID} bond can have an
 arbitrary shape and size, with the only constraint that its dimensions exceed
 5\,\mum\ by 5\,\mum. Additionally, the reduction of one process step is
 expected to lower the costs once the process is established in industry. In the
 interconnection-step the read-out chip and the sensor are brought together.
%
%
 In the bump-bonding process a built-in self alignment due to the surface
 tension of the bump-balls is exploited, while the {SLID} interconnection has to
 rely on the pick-and-place precision for the placement of the read-out chips on
 the handle wafer, when the technique is applied in the chip-to-wafer
 approach. If a high accuracy can be achieved in the pick-and-place procedure,
 the pitch of the {SLID} connections can be as low as approximately
 20\,\mum~\cite{Klumpp20mum}, which is not possible for the bump-bonding offered
 for industrial applications. In the final step, the actual bond is formed by
 pressing the two layers together. While a non functional bump-bonded module
 with a broken read-out chip or sensor can be separated again for repair by
 reheating this is not possible for {SLID} assemblies, due to the higher
 temperature needed.

 Other innovative interconnection technologies are presently investigated for
 pixel modules for use in high energy physics experiments.
 These are the Ziptronix Direct Bond Interconnect (DBI) oxide bonding and the
 copper thermo-compression~\cite{3DFermi}, both under evaluation at Fermilab,
 and also the copper pillar interconnects~\cite{LETI-copper}, offered by
 CEA-LETI~\cite{LETI}.
%
%
\section{Technical Aspects and Mechanical Properties}
\subsection{Influence of the SLID Process on Silicon Sensors}
 Since the SLID interconnection was only known to work with integrated circuit
 (IC) devices, a production of diodes subjected to the SLID metallization and
 temperature treatment was carried out. 
 Compared to IC devices, the performance of sensors that are usually made from
 high resistivity silicon, is much more sensitive to high leakage currents
 caused by a diffusion of copper atoms into the silicon bulk.
 In the {SLID} process, to prevent diffusion of copper into the silicon bulk a
 barrier layer of Titanium Tungsten (TiW) is needed.
 The diodes were used to verify the functionality of this TiW diffusion barrier
 with thin silicon sensors.

 To model both sides of the SLID metallization, two 6-inch wafers with various
 thin {p-in-n} diodes were produced. The sensor concept uses an SOI technology
 with an active sensor wafer that can be thinned to a desired thickness, and
 that is oxide bonded to a handle wafer for mechanical stability.
 The p-in-n option was chosen since no difference in the sensitivity of p-in-n
 and n-in-p sensors towards copper atoms is expected and the n-type wafers were
 easily procurable. The implemented diodes have an area of $10$\,mm$^2$ with
 different guard-ring designs and are thinned down with the HLL thinning
 technology to an active thickness of $50$\,\mum. Together with the handle
 wafer, the total thickness of the wafers is $500$\,\mum.
%
\begin{figure}[tbh]
\centering
\subfigure[]{
\includegraphics[width=0.20\textwidth]{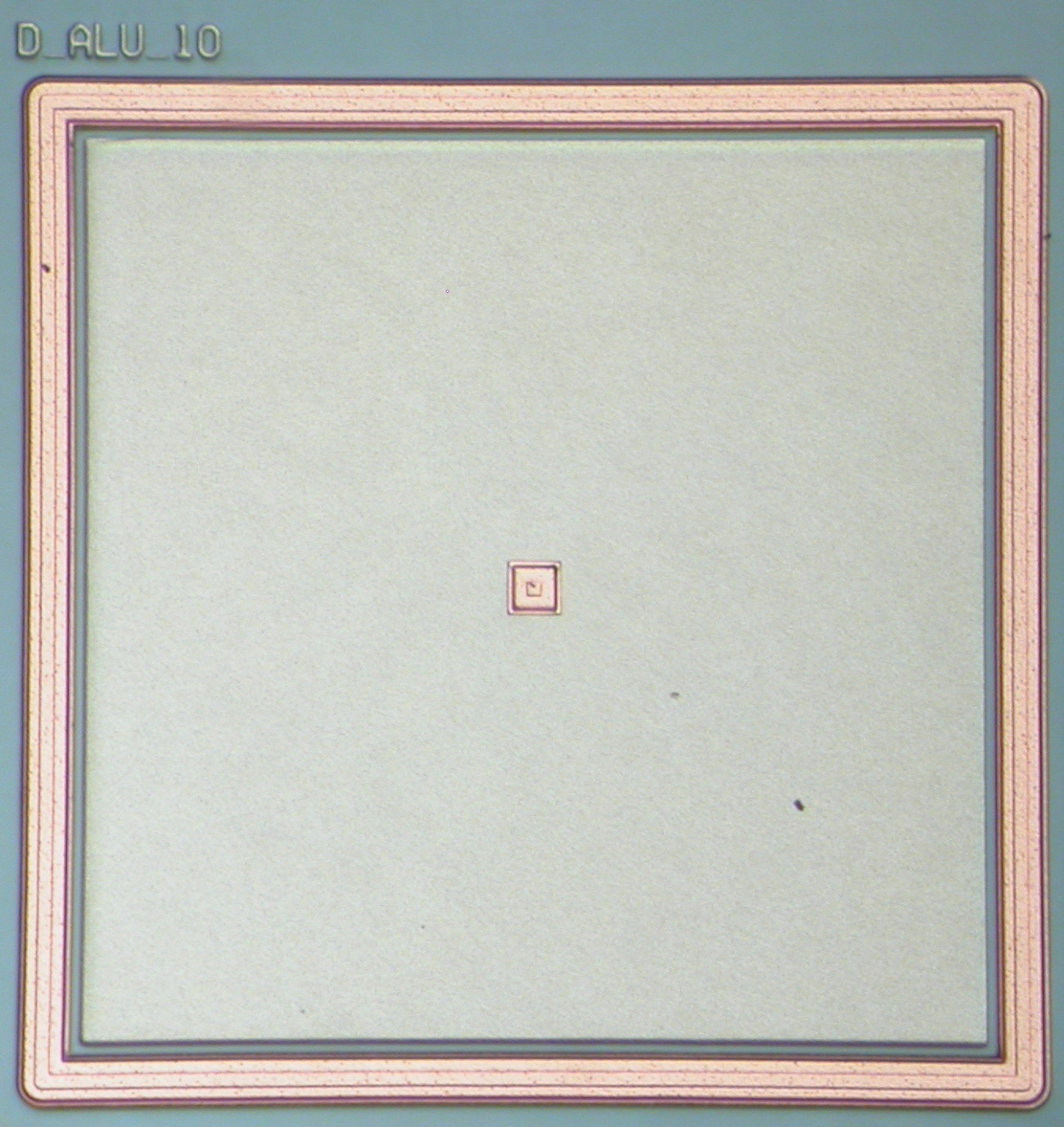}
\label{fig:CuDiode2}
}
\subfigure[]{
\includegraphics[width=0.20\textwidth]{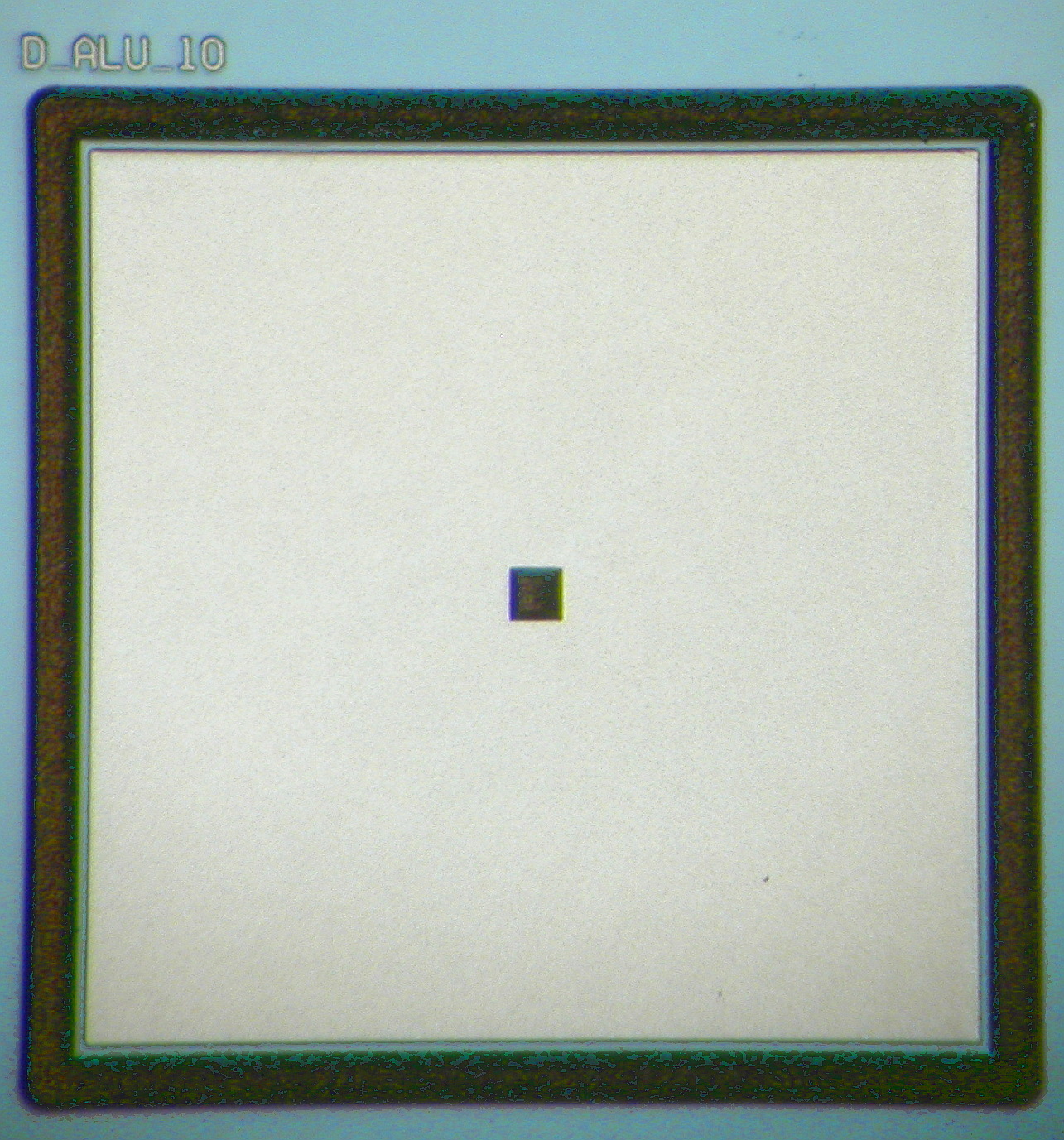}
\label{fig:CuSnDiode2}
}
\caption[Photographs of thin p-in-n diodes used to test the performance of the
  TiW diffusion barrier.]{Photographs of thin p-in-n diodes used to test the
  performance of the TiW diffusion barrier. The Cu can be seen as a pale orange
  surface while the Sn appears black under the coaxial illumination. (For
  interpretation of the references to colour in this figure caption, the reader
  is referred to the web version of this paper.)}
\label{fig:Diode2}
\end{figure}
%
\begin{figure*}[tbh]
\centering
\subfigure[]{
\includegraphics[width=0.47\textwidth]{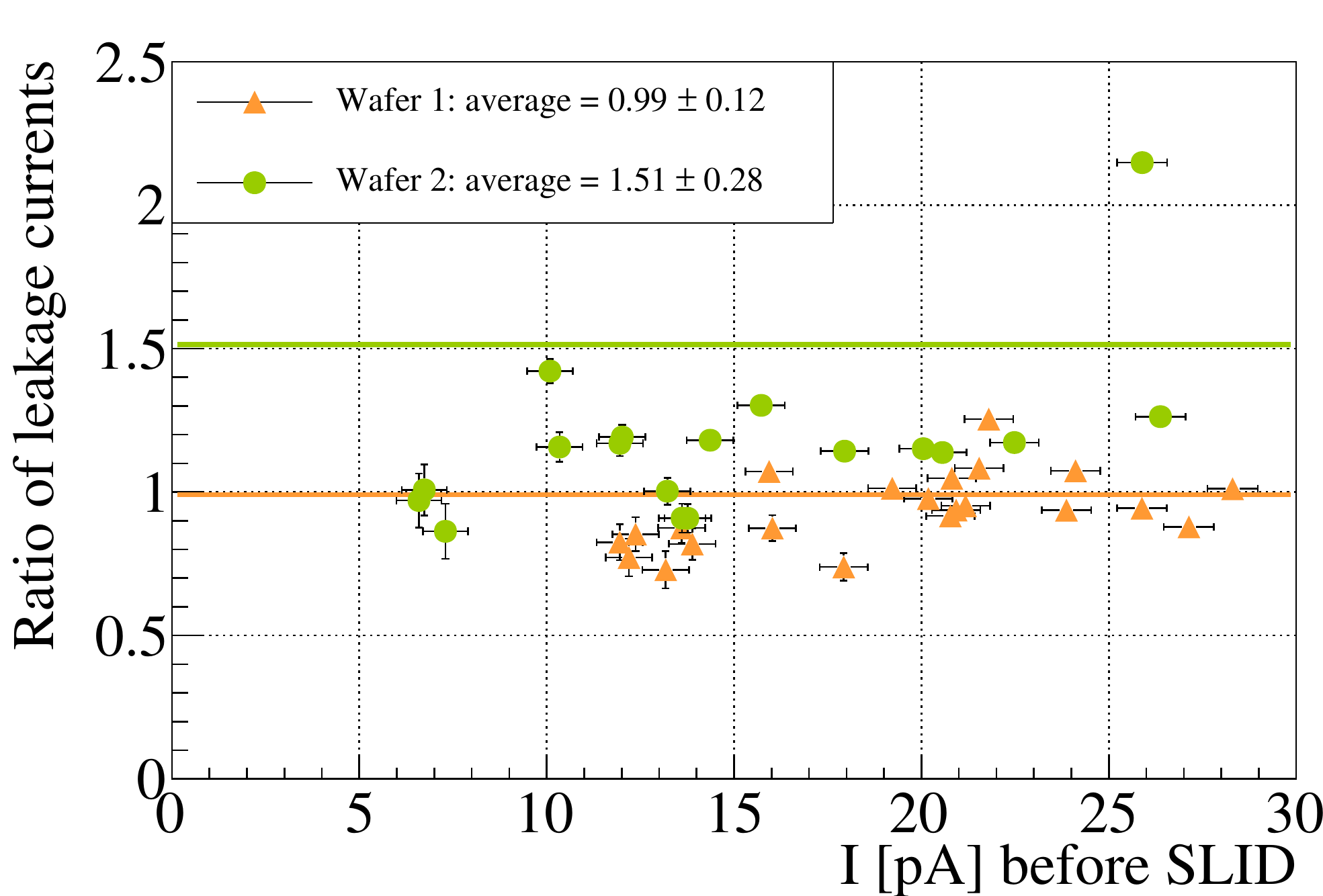}
\label{fig:aftercu}
}
\subfigure[]{
\includegraphics[width=0.47\textwidth]{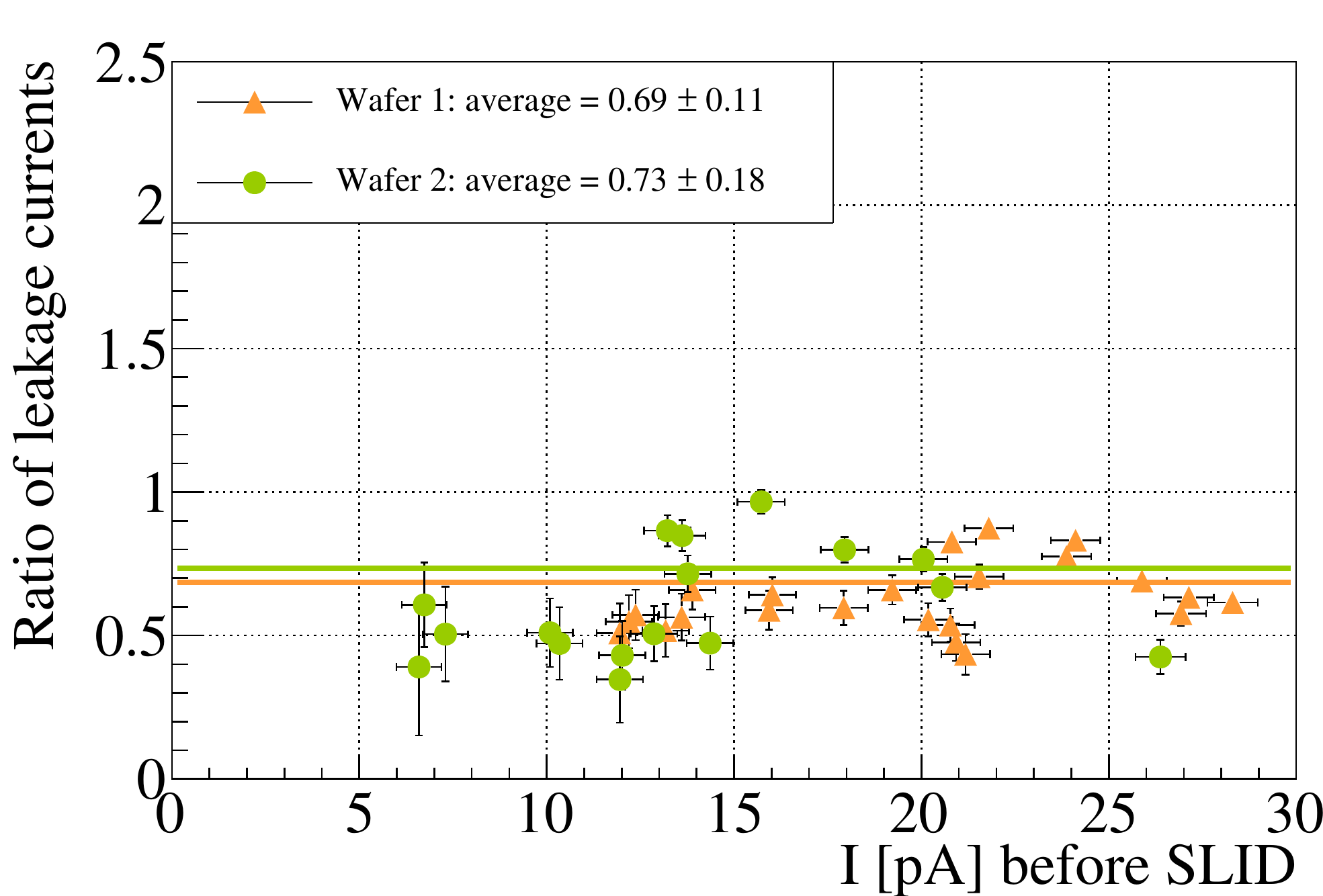}
\label{fig:afterheat}
}
\caption[Ratio of leakage currents after the application of the SLID
  metallization.]{Ratio of leakage currents of the thin p-in-n diodes with
  respect to the untreated diodes as a function of the leakage current. The
  ratios are shown in \subref{fig:aftercu} after the application of the SLID
  metallization and in \subref{fig:afterheat} after the SLID temperature
  treatment.}
\end{figure*}

 On both wafers shown in \fig~\ref{fig:Diode2}, a $100$\,nm thin layer of TiW
 was applied to the aluminium contact pads of all diodes, followed by an
 electroplating of Cu. For the first wafer, shown in \fig~\ref{fig:CuDiode2},
 the thickness of the Cu is around $1$\,\mum\ and no further layers are applied.
 The second wafer, which is displayed in \fig~\ref{fig:CuSnDiode2}, was equipped
 with $5$\,\mum\ of Cu and $1$\,\mum\ of Sn.
 Hence, both sides of the SLID metallization are replicated separately. The
 reduced thickness of the Cu layer on the first wafer of $1$\,\mum, compared to
 the $5$\,\mum\ used in the SLID interconnection, is assumed to be sufficient
 for an investigation of the full impact caused by a possible copper
 diffusion. For the second wafer, the thickness of the Sn was chosen to be a
 little less than half of the $3$\,\mum\ used in the SLID process. This ensures,
 that the Sn can be completely absorbed by the single Cu layer of the wafer.

 In a first step the leakage currents of the diodes were measured before the
 application of any SLID metal layers. The same diodes were measured after the
 application of the TiW and Cu for the first wafer and TiW, Cu, and Sn for the
 second wafer. 
 Shown in \fig~\ref{fig:aftercu} are the ratios of these leakage currents of the
 measured diodes of both wafers obtained after various production steps and at
 $50$\,V.
 The currents were determined by a linear fit to the plateau region of the
 leakage current characteristics. The uncertainties assigned are the combination
 of the one standard deviation uncertainties calculated from the measurement
 uncertainties of the Keithley-487 picoamperemeter~\cite{keithley} and the fit
 uncertainties.
 The leakage currents of the diodes on both wafers do not increase to a level
 that could be dangerous for the sensor operation. On wafer~1, the average
 current of the diodes at $50$\,V is unchanged while on wafer~2, it increases by
 about 51$\%$. 

 Removing the outlier (i.e.~the diode of wafer~2 in \fig~\ref{fig:aftercu} that
 has a ratio of leakage currents of about 2.2) from the analysis, which had a
 defect not related to the SLID interconnection, an average increase of only
 18$\%$ is found. These measurements show that during the application of the
 SLID metallization no copper diffuses into the sensor, since this would lead to
 an increase of the leakage current by several orders of
 magnitude~\cite{Istratov}.

 In a next step, both wafers were heated in the standard processing atmosphere
 to 320\,\degC\ for 15\,min to simulate the SLID temperature treatment, and to
 start the Solid-Liquid InterDiffusion of the Sn into the Cu. An actual
 connection of the wafers was not performed. After the temperature treatment the
 leakage currents of the diodes showed a slight decrease, as shown in
 \fig~\ref{fig:afterheat}. Compared to the measurements before any SLID
 processing steps, the currents are only 69$\%$ and 73$\%$ of the initial values
 for wafers~1 and 2, respectively.
 This is expected to be due to the annealing of defects in the silicon bulk
 caused by the applied temperature.
%
%
\subsection{Alignment Precision and Interconnection Efficiency of the SLID Interconnection}
 One of the key performance parameters to judge the applicability of the SLID
 interconnection technology is its connection efficiency $p$. It is defined as
 the probability that a given single SLID interconnection is successful. The
 corresponding inefficiency, i.\,e.\ the probability $p_{\rm not}=1-p$ of a
 fault of a given connection, is the figure of merit commonly used and given in
 the results below. To calculate the inefficiency from measurements of
 structures with a group of serial SLID interconnections, a binomial probability
 distribution is assumed. From the number $n$ of SLID interconnections per group
 and the fraction $P$ of groups with all connections working,
%
\begin{equation}
  p_{\rm not}=1-P^{1/n}
\end{equation}
%
 is derived. The inefficiency should be as low as possible; for example it was
 required to be smaller than $10^{-4}$ for the present ATLAS pixel
 modules~\cite{pixeltdr}.

\begin{figure*}[tbh]
  \begin{center}
  \subfigure[] {
    \label{fig:metaldummies:wafer}
    \includegraphics[width=0.51\textwidth]{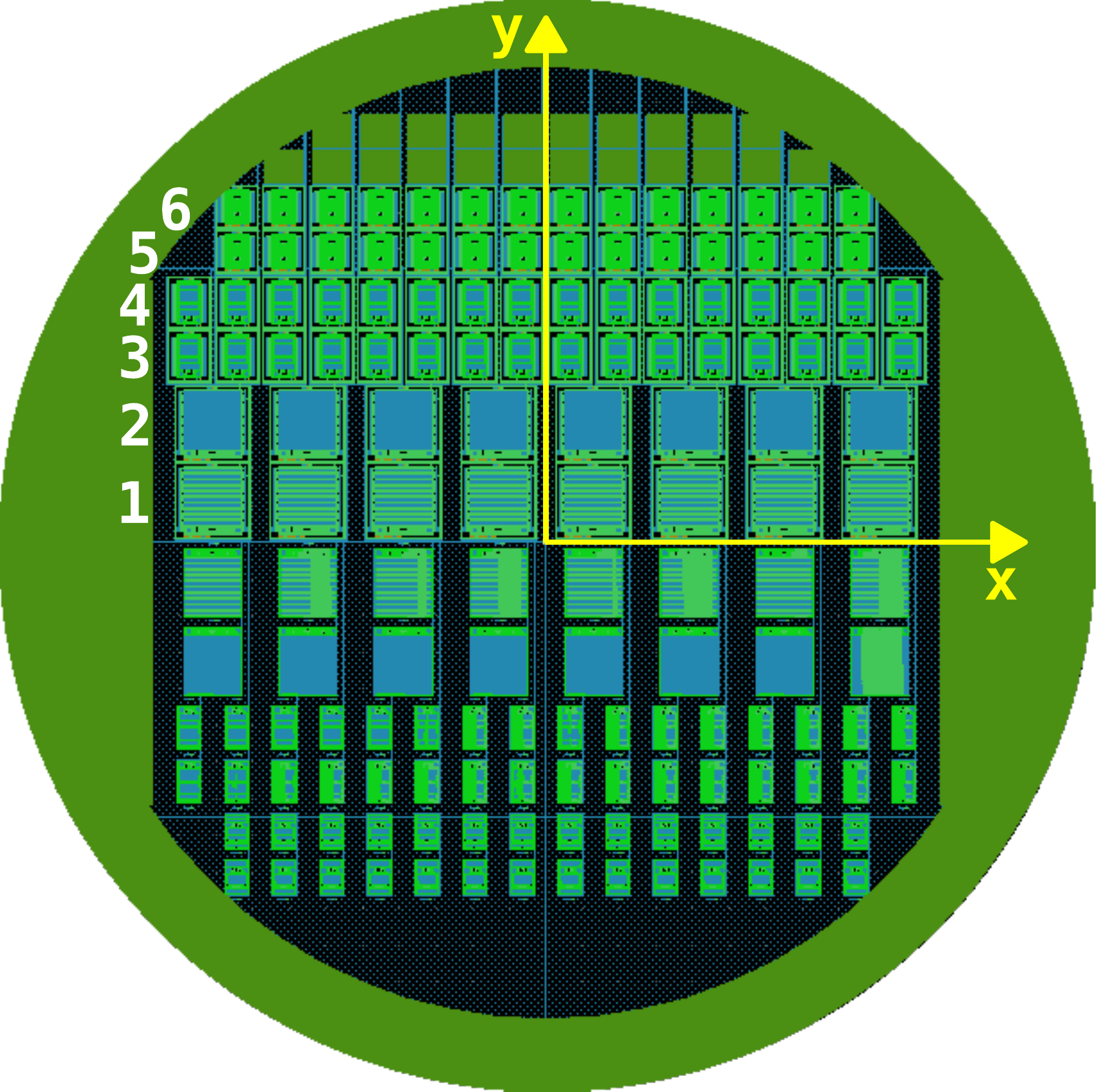}
  }
  \qquad
  \subfigure[] {
    \label{fig:metaldummies:chain}
    \raisebox{3mm}{\includegraphics[width=0.45\textwidth]{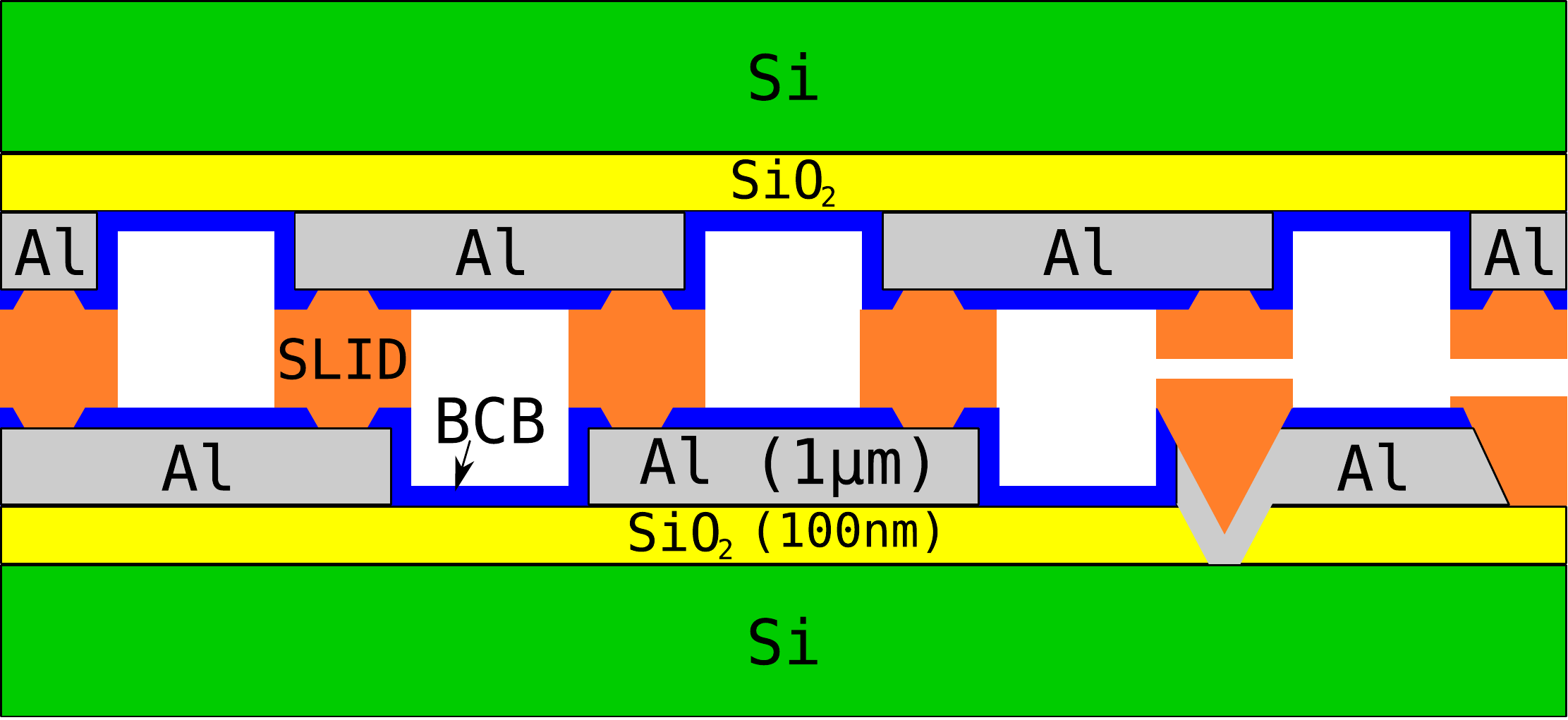}}
  }
  \qquad
  \subfigure[] {
    \label{fig:metaldummies:alignment}
    \includegraphics[width=0.25\textwidth]{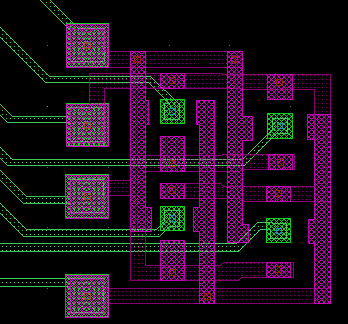}
  }
  \caption{Overview and detailed schematics of the SLID prototype production. In
    \subref{fig:metaldummies:wafer} the wafer map of the SLID dummy devices is
    shown, in \subref{fig:metaldummies:chain} the schematics of an individual
    daisy chain is displayed, and finally, \subref{fig:metaldummies:alignment}
    shows an electrical structure to verify the mechanical alignment.  (For
    interpretation of the references to colour in this figure caption, the
    reader is referred to the web version of this paper.)
  \label{fig:metaldummies}}
  \end{center}
\end{figure*}

 To measure $p_{\mathrm{not}}$ of the SLID interconnection, its dependence on
 the alignment precision, and the sensitivity to disturbances of the device
 planarity, a SLID prototype production was carried out.
 For this, a 6-in.-wafer layout designed at the MPP, and shown in
 \fig~\ref{fig:metaldummies:wafer}, was used. The layout includes a total of 152
 test devices, which rely solely on structured metallisation directly on the
 SiO$_2$ but do not have any implants. The SLID contact positions are symmetric
 with respect to the $x$-axis and hence, two of these wafers can be connected by
 rotating one around its symmetry axis by $180^\circ$ and placing it onto the
 other.
 Through this, the 76 devices in the northern half of one wafer, which are
 referred to as sensor devices, are connected to the 76 chip devices of the
 southern part of the other wafer.
%

 A large fraction of the area of each device is filled with daisy chains which
 are a serial wiring scheme of a large group of SLID interconnections in a row
 with alternating aluminium traces on the sensor- and chip-side, as shown in
 \fig~\ref{fig:metaldummies:chain}. If a potential difference is applied to the
 ends of a daisy chain, a current can only flow provided all SLID
 interconnections are functional. Hence, a large number of SLID interconnections
 are tested at the same time. The daisy chains of the 76 devices of the sensor
 side are equipped with aluminium traces leading to contact pads for needle
 probes at the ends of the daisy chains. On the chip devices, there are no
 traces since this part of the devices is cut off during the singularisation to
 enable access to the contact pads of the sensor devices. Hence, after
 connecting two wafers only those 76 structures can be used where the sensor
 devices are on the lower wafer which is not cut.

 Rows~1 and 2 of larger devices above the horizontal wafer axis contain daisy
 chains which have the same geometry as an ATLAS pixel sensor. This means that
 the metal traces occupy the same areas as the pixel implants do in the
 sensors. In addition, aluminium lines are implemented to connect every second
 pair of traces to form an open chain. The SLID interconnection of open chains
 from chip and sensor devices leads to closed, i.\,e.~conducting chains, as
 shown in \fig~\ref{fig:metaldummies:chain}. In row~1, the SLID pad size is
 $27\times58$\,\mumq\ with a small pitch of 50\,\mum\ and a large pitch of
 400\,\mum. This corresponds to the SLID pad dimensions used for the prototype
 pixel modules discussed below. The chains of row~2 have identical pitches but
 the SLID pads are of similar size as the n-type implants in the ATLAS pixel
 sensors, i.\,e.~$27\times 360$\,\mumq. Within the smaller devices in rows~3 to
 6 of the wafer, a variety of SLID pad sizes and pitches are implemented in
 different daisy chains. They range from $30\times30$\,\mumq\ with a pitch of
 $60$\,\mum\ to $80\times80$\,\mumq\ with a pitch of 115\,\mum\ as detailed in
 \tab~\ref{tab:SLID}.
%
\begin{table}[tbp]
  \begin{center}
   \begin{tabular}{|c|c|c|c|c|}
    \hline
    Pad size  & Pitch   & Aplanarity & Connections & Inefficiency  \\
    $[\rm \mu m^2]$& $[\rm \mu m]$& $[\rm \mu m]$& measured &
    $p_{\rm not}[10^{-3}]$\\
    \hline
    \hline
    $30\times 30$ & $60$ & $-$ & $8288$ & $<0.36$  \\
    $80\times 80$ & $115$ & $-$ & $1120$ & $<2.7$  \\
    $80\times 80$ & $100$ & $-$ & $1288$ & $<2.3$  \\
    $27\times 60$ & $50,400$ & $-$ & $24160$ & $0.5\pm0.1$ \\
    $30\times 30$ & $60$ & $0.1$ & $5400$ & $1.0\pm0.4$ \\
    $30\times 30$ & $60$ & $1.0$ & $5400$ & $0.4\pm0.3$ \\
    \hline
    \end{tabular}
    \caption{Geometrical parameters and performance of various SLID
      interconnection options.
      \label{tab:SLID}}
\end{center}
\end{table}
%
 In addition, in rows~3 and 4 special chains are implemented which have a part,
 where deliberately either the SiO$_2$ or the aluminium layer is missing. This
 leads to a lowering of the SLID pads by 100\,nm or 1\,\mum\ as illustrated on
 the right side of \fig~\ref{fig:metaldummies:chain}. With these degradations of
 the device planarity the sensitivity of the SLID interconnection to surface
 imperfections is investigated.

 Furthermore, electrical and optical alignment structures are introduced in the
 devices. The electrical alignment structures consist of SLID pads that are only
 connected if the devices are misaligned. A section of the wafer map containing
 one of the alignment structures which measures a misalignment of
 (2.5--15)\,\mum\ is shown in \fig~\ref{fig:metaldummies:alignment}.
 The structures shown in green are located on the sensor wafer and consist of
 eight metallized squares, four small ones and four larger ones. The structures
 drawn in red are located on the chip side.
 Depending on the size of the misalignment different counterpart pads match, and
 are electrically conducting, which is verified with probe needles on external
 pads (not shown).
 In the presented case of perfect alignment, the four large square contacts of
 both sides are connected, while the small green square contacts have no counter
 part on the chip. 
 If, as an example, a misalignment of 3\,\mum\ is introduced, the lower left
 SLID pad in \fig\,\ref{fig:metaldummies:alignment} can contact one or two of
 the surrounding red structures on the chip.
 This forms a conducting channel which can be identified by contacting the
 corresponding probe pads. Since the sensor side squares will connect to
 different counterparts, not only the magnitude but also the direction of the
 misalignment can be identified.
 Further alignment structures on each device allow for measuring a misalignment
 of up to 30\,\mum.

\begin{figure}[h]
  \begin{center}
    \includegraphics[width=0.47\textwidth]{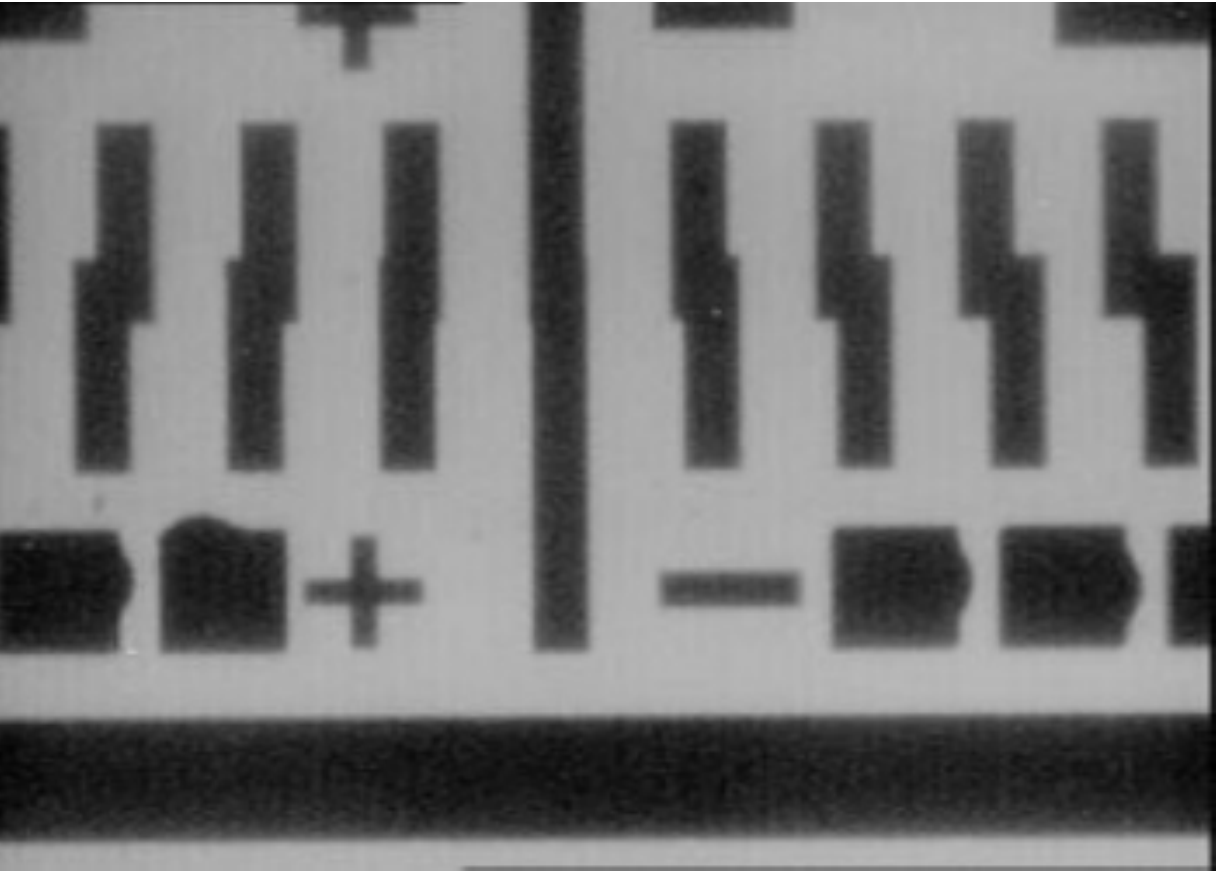}
    \caption[Optical alignment of the SLID interconnections.]{Infra-red image of
      a well aligned vertical alignment vernier scale. A perfect alignment is
      reached if only the central long aluminium lines of both wafers completely
      overlap. For each 6\,\mum\ of misalignment, the lines that completely
      overlap are shifted by one to the left or right.}
    \label{fig:nonius}
  \end{center}
\end{figure}

 The optical alignment structures are aluminium vernier scales that are
 implemented partly on the sensor- and partly on the chip side of the packages,
 as shown in \fig~\ref{fig:nonius}.
 Using an infra-red microscope they allow to determine the relative misalignment
 with an accuracy of 3\,\mum.
 The measurements of the SLID daisy chains were carried out with a
 Keithley-6517A electrometer~\cite{keithley}, supplying a small voltage to the
 ends of the chains and measuring the current. Through this, also the resistance
 of the chains are measured and a mean resistance per SLID connection is
 determined.  Using an infra-red microscope the relative misalignment is
 determined.
%
%
\subsubsection{Wafer-to-Wafer Interconnection}
 The wafers were interconnected in a wafer-to-wafer approach. For the majority
 of the chains all SLID connections were functioning resulting in finite
 resistances ranging from $(0.25\pm0.12)\rm\,\Omega$ to $(1.5\pm1.7)\rm\,\Omega$
 per SLID connection, where the uncertainties are the one standard deviations of
 the measurements from various equivalent chains. The chain resistances do not
 directly correlate to the size of the SLID pads but rather to the number of
 SLID connections per row ranging from 46 to 302 connections. This leads to the
 conclusion that the dominating contribution to the resistance is not caused by
 the SLID metal layers, but rather by the contact between them and the aluminium
 traces. 
 This contact is made underneath each SLID pad by creating a circular opening
 with 10\,\mum\ nominal diameter in the BCB passivation layer covering the whole
 wafer, displayed in \fig~\ref{fig:metaldummies:chain}. The openings have the
 same diameter for all pads of all chains.

 \tab~\ref{tab:SLID} summarizes the results of all daisy chain measurements and
 includes the total number of SLID connections tested.
 The SLID inefficiency is less than $10^{-3}$ for most of the chain types
 without a deliberately introduced aplanarity.
 The exception are the structures of row~1 for which 24160 contacts were
 measured, and for which 10 out of 80 chains with 302 connections each were
 interrupted.
 In those cases, where no interrupted contacts were found, an upper limit at a
 90$\%$ confidence level is reported. The smallest limit observed is $4\cdot
 10^{-4}$, consequently higher statistics data are needed to verify that an
 inefficiency of less than $10^{-4}$ is met by the process.
 Some chains have been produced without the aluminium layer below the SLID pads.
 This is an exaggerated situation that is much more severe than the typical
 thickness variations of the aluminium layer, and does not occur in real
 applications.
 Even those chains result in a connection inefficiency per pad of
 $(0.4\pm0.3)\cdot 10^{-3}$, clearly showing that the SLID interconnection is
 not severely affected by variations of the surface planarity up to 1\,\mum.
%
%
 
 The optical inspections of the vernier scales as well as the measurements of
 the electrical alignment structures showed a very good alignment accuracy of
 better than 5\,\mum\ for the first and about (5--10)\,\mum\ for the second pair
 of interconnected wafers.
%
\subsubsection{Chip-to-Wafer Interconnection}
 In another prototype run ATLAS FE-I3 read-out chips were interconnected to
 fully functional thin pixel sensors. The sensors were produced on p-bulk {FZ}
 wafer using the {MPG-HLL} thinning process with a final active thickness
 \dactive\ of 75\,\mum. The specific resistivity of these wafers is $\rho\geq
 2$\,k$\Omega$cm. A discussion of the electrical characteristics of all
 structures within this production can be found in~\cite{BeimfordePhD}. The full
 depletion voltages $V_{\mathrm{fd}}$ were found to be ($30\pm5$)\,V, with the
 exception of one pixel device that did not reach a plateau in the leakage
 current, i.\,e.\ where the breakdown voltage $V_{\mathrm{bd}}$ was lower than
 $V_{\mathrm{fd}}$. This corresponds to a yield of $79/80\approx98.8\%$.
%
\begin{figure}[t!bh]
\centering
\includegraphics[width=\columnwidth]{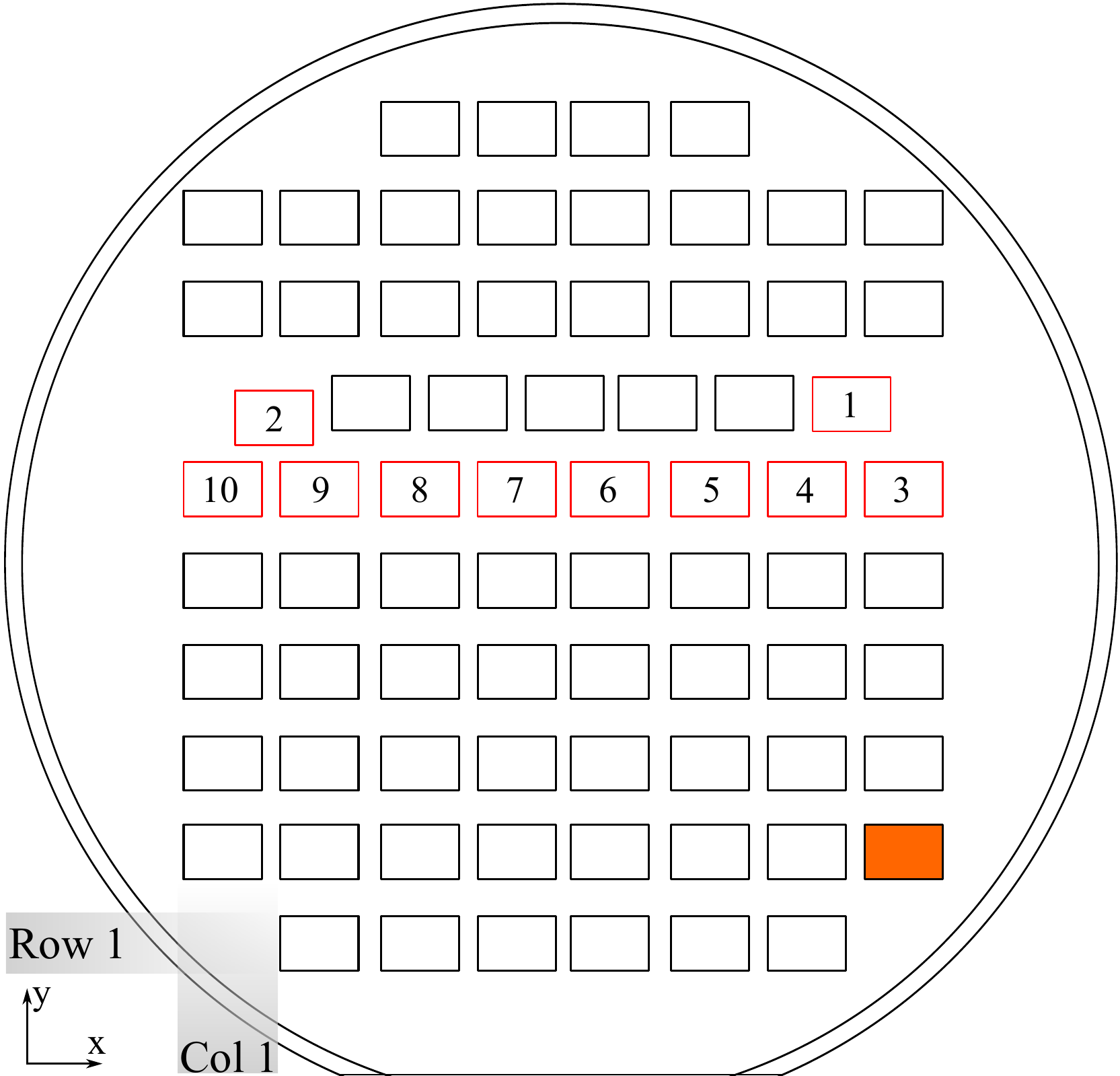}
\caption{Population layout of the handle wafer. Read-out chips corresponding to
  a compatible sensor in the wafer layout are indicated in red. The given
  numbers indicate the modules for further reference. Dummy read-out chips for
  mechanical stability are drawn in black. The read-out chip indicated in orange
  is missing. The coordinate system referred to in the following is also
  indicated. (For interpretation of the references to colour in this figure
  caption, the reader is referred to the web version of this paper.)}
\label{fig:SLID_HandleWafer}
\end{figure}
%
 For the 79 pixel devices, an over depletion $V_{\mathrm{bd}}/V_{\mathrm{fd}}$
 of $3.7\pm1.0$ up to $15\pm2$ can be reached. Finally, their leakage currents
 in the plateau region were determined to be below 10\,nA/cm$^{2}$.

 For the interconnection, operating a flip-chipping machine in a pick and place
 mode, in a chip-to-wafer process, a handle wafer was populated with known
 working read-out chips at the positions of compatible pixel structures on the
 sensor wafer side as indicated by the red numbered rectangles in
 \fig~\ref{fig:SLID_HandleWafer}.
 Due to the high applied pressure in the process, and to achieve good precision
 in the pick and place process, a regular pattern of chips on the handle wafer
 is mandatory. Consequently, the rest of the handle wafer was populated
 regularly with read-out chips (indicated in black). The electroplated {SLID}
 pad structure is the same for the working and for the dummy read-out chips.
 An excellent alignment of the read-out chips on the handle wafer with respect
 to their nominal positions, known from the design of the sensor wafer, is needed,
 given the small pitch and SLID pad sizes in combination with the needed minimal
 overlap of $5\,\mum\,\times\,5\,\mum$. 

 Additionally, it is important that rotations of the read-out chips are below
 about 0.5$^{\circ}$. Although, a global misalignment can be corrected for by
 adjusting the relative position of the two wafers in the wafer-to-wafer
 interconnection process, these requirements demand cutting-edge pick-and-place
 technology.
%
\begin{figure}[t!bh]
\centering
\subfigure[]{
\includegraphics[width=0.53\columnwidth]{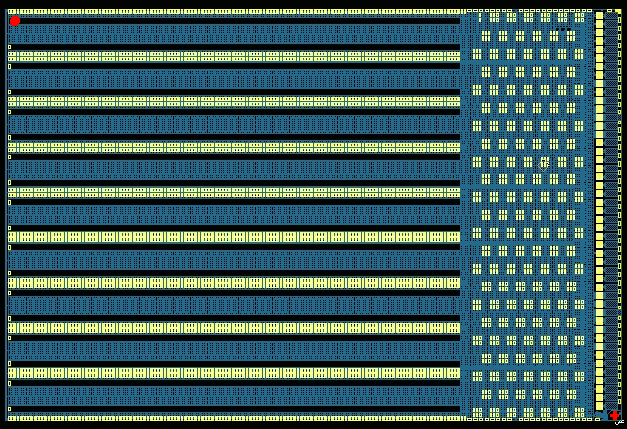}
\label{fig:SLIDAlignCross_sketch}
}
\subfigure[]{
\includegraphics[width=0.39\columnwidth]{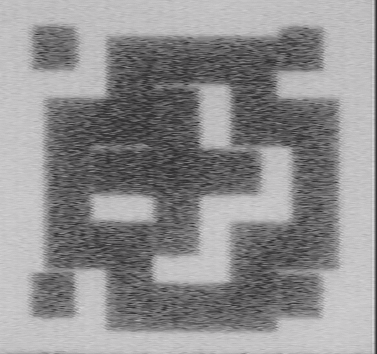}
\label{fig:SLIDAlignCross}
}
\caption[{SLID} pad distribution over the FE-I3 read-out chip and alignment
  marks for the interconnection.]{\subref{fig:SLIDAlignCross_sketch} {SLID} pad
  distribution over the FE-I3 read-out chip (yellow rectangles). The alignment
  marks are indicated in red. The pad is in the upper left corner, the cross is
  in the lower right corner. \subref{fig:SLIDAlignCross} Infra-red image of an
  alignment cross after interconnection. The cross has a total dimension of
  150\,\mum\ in both directions. (For interpretation of the references to colour
  in this figure caption, the reader is referred to the web version of this
  paper.)}
\label{fig:SLID_AlignmentMarks}
\end{figure}

 The positions of the alignment marks (cross and circle) are indicated in red in
 \fig~\ref{fig:SLIDAlignCross_sketch}. In \fig~\ref{fig:SLIDAlignCross}, an
 infra-red picture of a cross alignment mark is depicted for a connected
 stack. Based on these images, the quality of the alignment was determined after
 interconnection.
%
\begin{table}[htb]
\begin{center}
\begin{tabular}{|l|r|r|r|l|}
\hline
Module&$\Delta x$ [$\mu$m]&$\Delta y$ [$\mu$m]&Tilt [$^{\circ}$]&Connected [$\%$]\\
\hline\hline
1     & $-6$     & $-22$   & $-0.25$           & 100               \\  
2     & $-139$   & $-40$   & $-0.25$           &                   \\  
3     & $-23$    & $-34$   & $-0.38$           & 100               \\  
4     & 44       & 73      & 0.72              &                   \\  
5     & $-34$    & $-58$   & $-0.61$           &                   \\
6     & $-8$     & $-19$   & $-0.21$           & $70.1^{+0.3}_{-0.3}$ \\  
7     & $-16$    & $-18$   & $-0.21$           & $66.5^{+0.4}_{-0.4}$ \\  
8     & $-17$    & $-25$   & $-0.23$           & $88.7^{+0.3}_{-0.5}$ \\  
9     & $-17$    & $-21$   & $-0.24$           & $94.5^{+0.1}_{-0.3}$ \\  
10    & $-16$    & $-25$   & $-0.26$           & 100               \\  
\hline
\end{tabular}
\caption{Residual misalignment of the alignment cross for the interconnected
  modules and the fraction of connected pixel cells. For the definition of
  connected and the evaluation of uncertainties, please refer to the
  text. Assemblies 2, 4 and 5 were not investigated due to their misalignment.
 \label{tab:Hot2alignmentfinal}}
\end{center}
\end{table}
%
 The residual misalignment after interconnection is summarised in
 \tab~\ref{tab:Hot2alignmentfinal}. In total, seven out of ten assemblies were
 built successfully, i.\,e.\ without shorts or open connections caused by
 misalignment. For the assemblies 2, 4 and 5 the misalignment is too large for
 the pixel assemblies to be functional.
 To improve the precision of the alignment for future productions, a new and
 more precise pick-and-place machine will be employed. Additionally, the
 possibility to exploit self alignment via evaporative liquid glues while
 populating the handle wafer is currently investigated at the
 EMFT~\cite{EMFTpriv}.

 Open connections were identified with a high statistics radioactive source
 measurement in which not connected pixel cells exhibit a low hit rate, because
 they can only contribute via electronic noise, but not via genuine signal. For
 the used statistics, and in the centre of the beam spot, around 150 hits per
 pixel are expected.
 A pixel cell is defined as connected, if it exhibits more than 50
 hits. Uncertainties are assessed by varying this threshold by $\pm10\%$. The
 percentages of connected pixel cells per module are summarised in
 \tab~\ref{tab:Hot2alignmentfinal}. While for module 1, 3 and 10 all pixel-cells
 are connected, module 6 exhibits around 30$\%$ of not connected pixel cells. A
 trend of the fraction of not connected cells to rise towards the centre of the
 wafer is found.

\begin{figure}[t!]
\centering
\subfigure[]{
\includegraphics[width=0.22\textwidth]{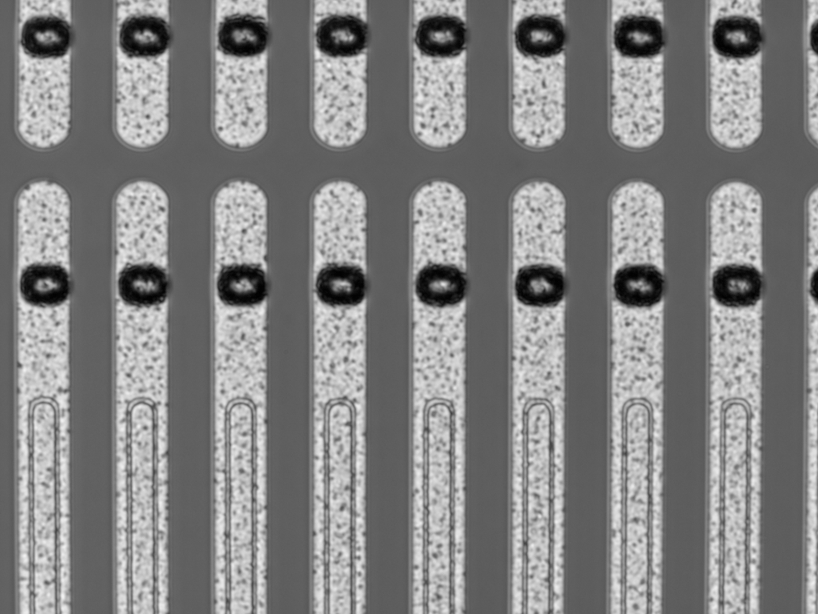}
\label{fig:SLID_not_connected_BCB_bad}
}
\subfigure[]{
\includegraphics[width=0.22\textwidth]{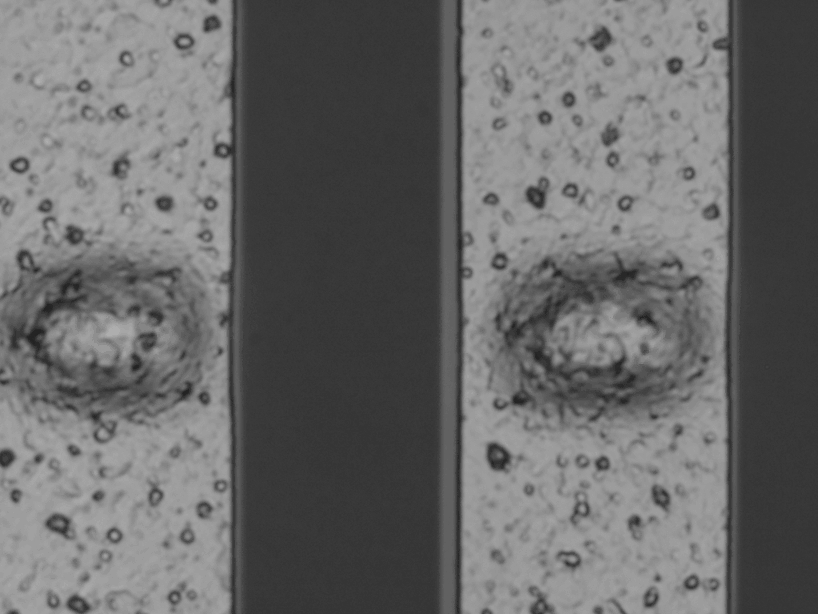}
\label{fig:SLID_not_connected_BCB_bad_zoom}
}
\subfigure[]{
\includegraphics[width=0.22\textwidth]{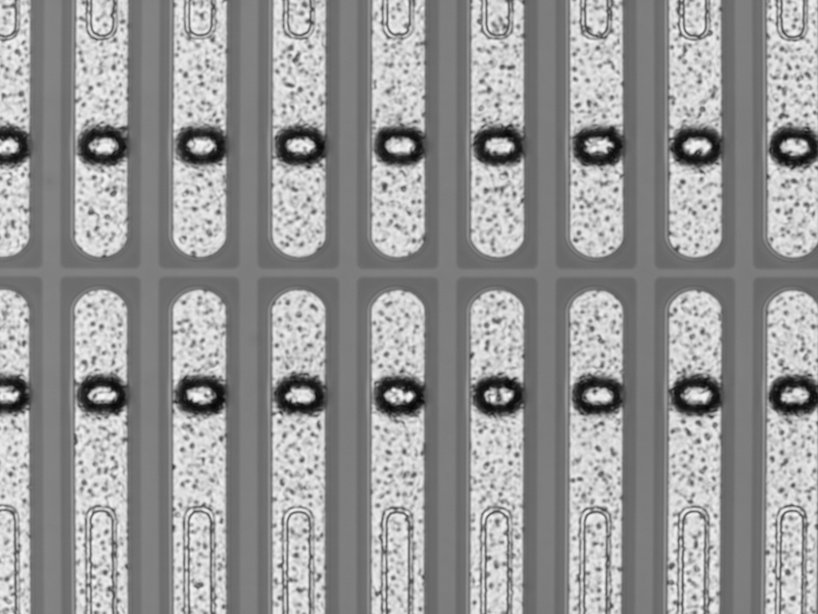}
\label{fig:SLID_not_connected_BCB_good}
}
\subfigure[]{
\includegraphics[width=0.22\textwidth]{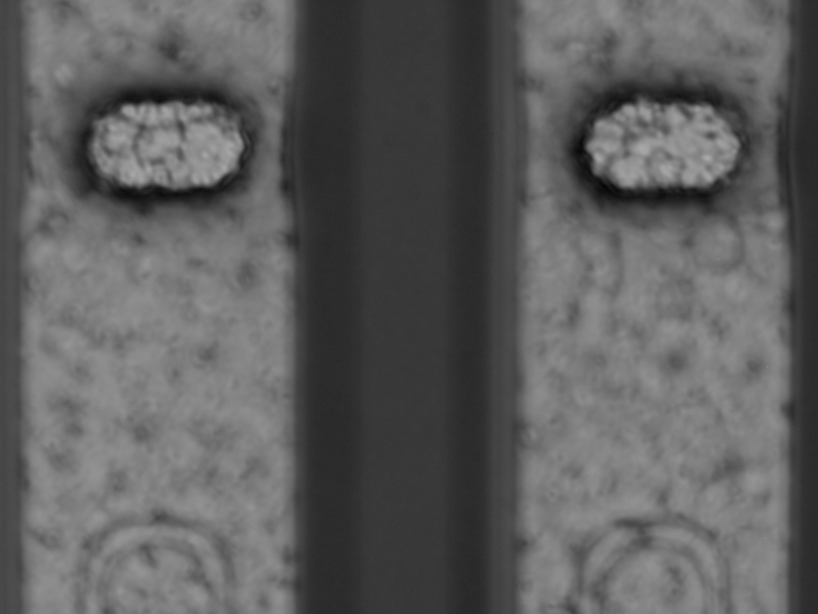}
\label{fig:SLID_not_connected_BCB_good_zoom}
}
\caption[Photographs of {BCB} passivation layers in the position corresponding
  to the {SLID} pads.]{Photographs of (a, b) an insufficiently opened {BCB}
  passivation layer in the position corresponding to the {SLID} pads, and (c, d)
  a fully opened {BCB} passivation layer. The horizontal distance between two
  openings is 50\,\mum\ in all photographs. 
 The BCB openings have an elliptic form with nominal lengths of 15\,\mum\ and
 22\,\mum\ for the two axes.}
\label{fig:SLID_not_connected_BCB}
\end{figure}

 Subsequent optical re-inspections of not yet connected sensor wafers from the
 same production revealed that the cause for these not connected pixel cells are
 imperfect openings of the BCB passivation layer underneath the SLID pads that
 show a radial trend across the wafer similar to the one observed for the not
 connected cells.
 Photographs of such not fully opened layers are depicted in
 \fig~\ref{fig:SLID_not_connected_BCB_bad} and
 \fig~\ref{fig:SLID_not_connected_BCB_bad_zoom}. For future module assemblies, a
 removal of residual {BCB} in the openings using an SF$_{6}$ plasma descum
 process offered by the Fraunhofer IZM~\cite{IZM} was investigated. In the
 optical inspection after the treatment all BCB contacts were found to be fully
 opened. \fig~\ref{fig:SLID_not_connected_BCB_good} and
 \fig~\ref{fig:SLID_not_connected_BCB_good_zoom} are photographs of fully opened
 contacts. Thus, this is not an issue for future productions.
%

 Another crucial factor is the stability of the connections in experimental
 conditions, where, in addition to high radiation levels, temperature cycles are
 present. Within the laboratory and during beam test measurements for all
 modules the numbers of not connected pixel cells did not change with numerous
 thermal cycles between 20\,\degC\ and $-50$\,\degC. Furthermore, no changes
 after irradiation up to a fluence of $10^{16}$\,\neqcm\ were observed. This is
 a strong indication that {SLID} interconnections are radiation hard and
 withstand thermal cycles.
%
%
\subsection{Mechanical Strength}
%
\begin{figure}[tbh]
\centering
\includegraphics[width=0.29\textwidth]{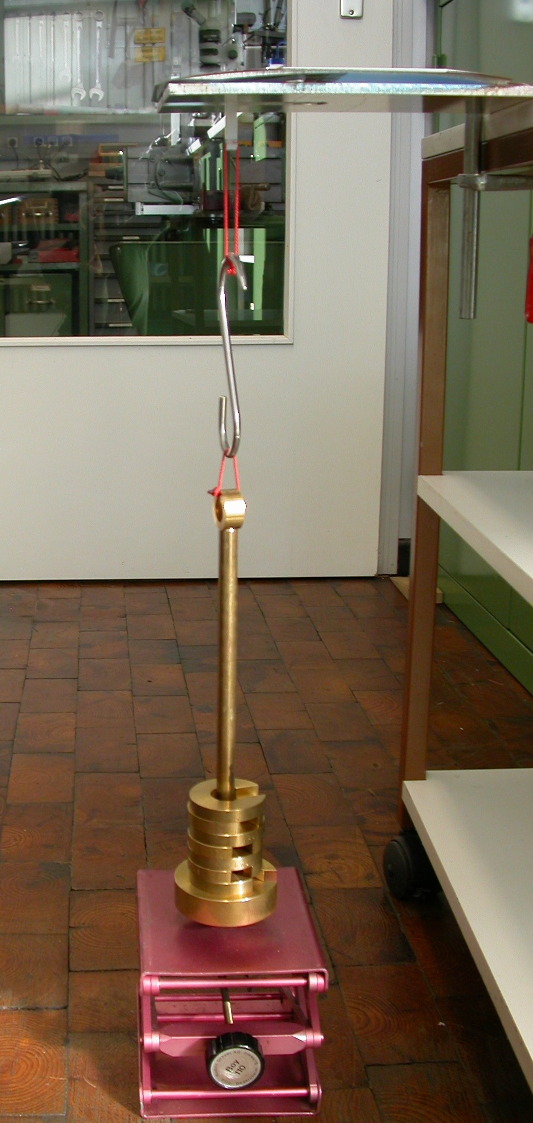}
\caption[The setup to determine and results of the mechanical strength
  test.]{Photograph of the mechanical strength test
  setup. \label{fig:SLID_Pulltest_setup}}
\end{figure}
%
\begin{figure}[tbh]
\centering
\includegraphics[width=0.50\textwidth]{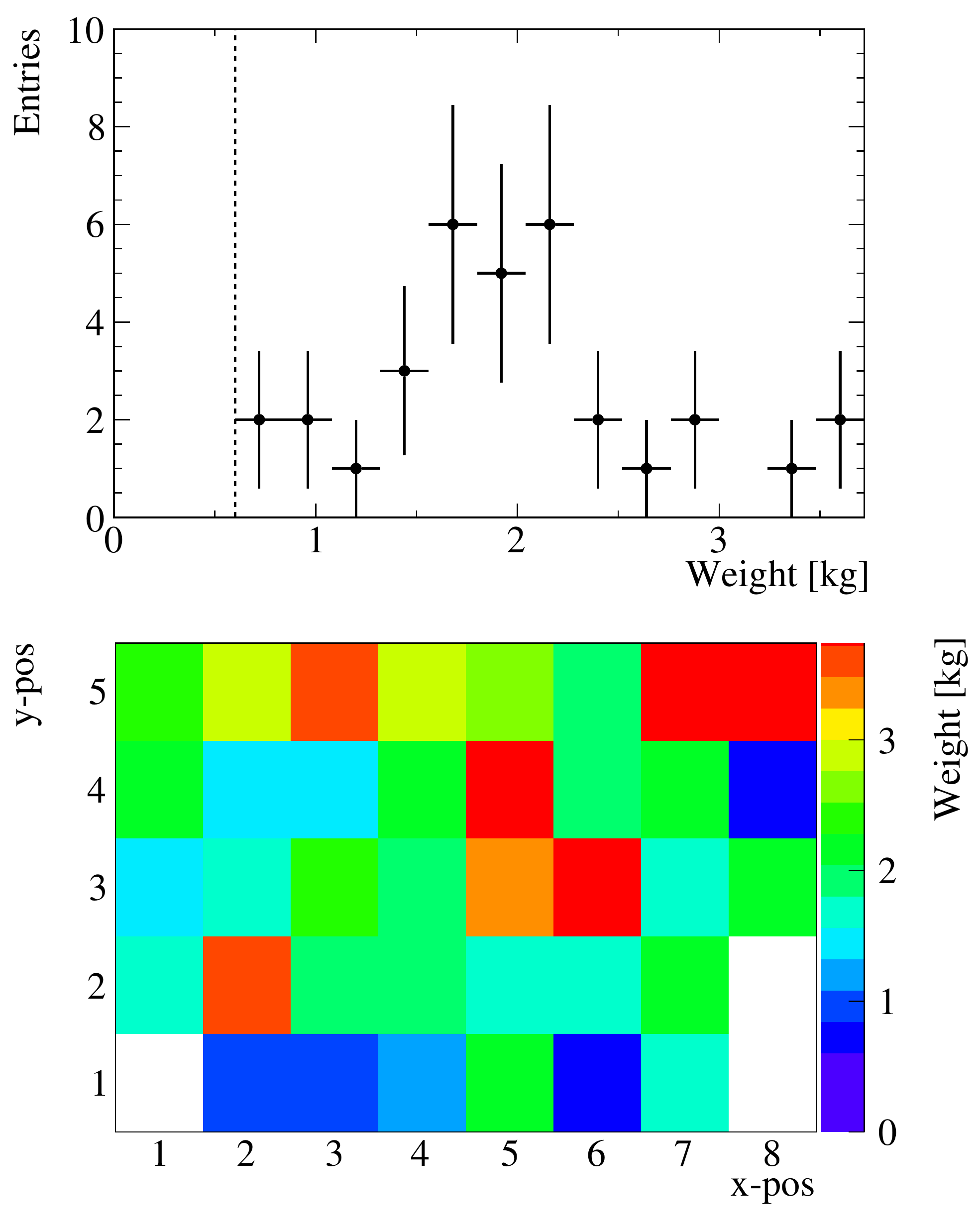}
\caption{The weight at which the read-out chip was separated from the sensor. In
  the top figure the distribution is shown. The dotted line indicates the
  minimal weight applied. In the lower figure the position of the read-out chip
  on the wafer is indicated. The read-out chip at position (8,2) fell off before
  the test; the positions (1,1) and (8,1) are not populated by design, as shown
  in \fig~\protect\ref{fig:SLID_HandleWafer}.
  \label{fig:SLID_Pulltest_result}}
\end{figure}

 A high mechanical strength is desirable for an interconnection technology, as
 it eases the handling of the device, ensures that bonds do not break
 accidentally, and that they are stable in time. To determine the mechanical
 strength, a piece of plexiglass was glued onto each dummy read-out chip (black
 in \fig~\ref{fig:SLID_HandleWafer}) in the lower half of the handle
 wafer. Subsequently, weight was hanged onto the plexiglass holder while the
 sensor wafer was stabilised in its position by a plexiglass support covering
 the full area except for the region around the read-out chip under study. After
 each increase of weight the strain was relieved using a small hoisting platform
 to apply the force in a controlled manner. Before adding the next weight the
 hoisting platform was lifted again. A photograph of the setup is depicted in
 \fig~\ref{fig:SLID_Pulltest_setup}. Due to the construction, the minimum weight
 applied is 0.6\,kg.

 The distribution of the weight needed to break the connection between sensor
 and read-out chip is given in \fig~\ref{fig:SLID_Pulltest_result}. No
 systematic trend across the wafer is appreciable and the weight needed for
 breakage is approximately two kilograms, which corresponds to 0.01\,N per
 {SLID} connection. This is of the same order of magnitude to what is found for
 other interconnection
 technologies~\cite{BumpGoPatent,Broennimann2006303,Eldring,Broennimann:gf0003,Cheah}. With
 the exception of extreme cases of misalignment, no significant correlation
 between the misalignment and the connection is found.

\begin{figure}[h!tb]
\centering
\subfigure[]{
\includegraphics[width=0.47\columnwidth]{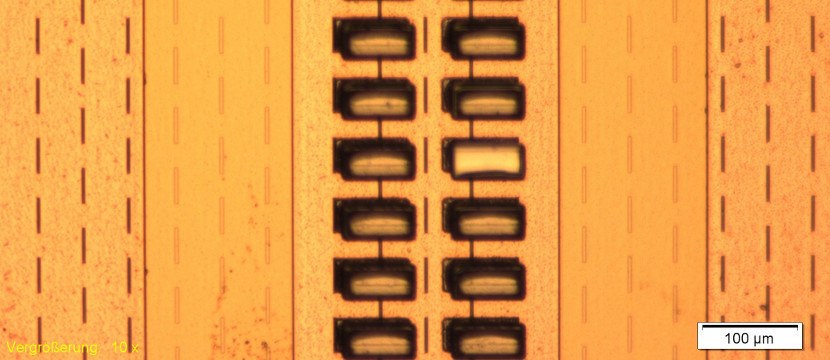}
\label{fig:100mu}
}
\subfigure[]{
\includegraphics[width=0.47\columnwidth]{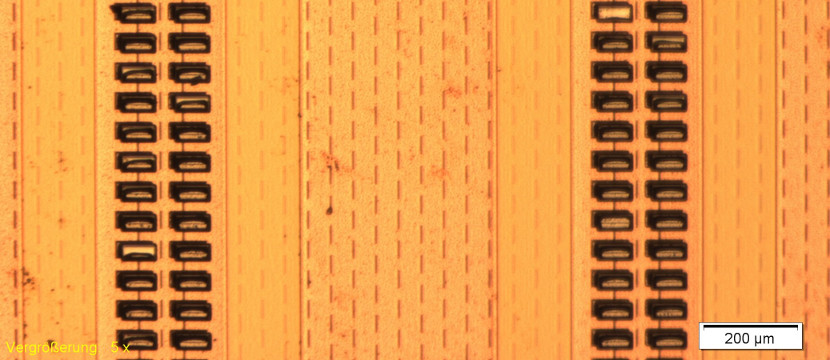}
\label{fig:200mu}
}
\caption[{SLID} stacks in the centre of the structure for a dummy read-out chip
  separated from the dummy sensor.]{Well aligned {SLID} stacks in the centre of
  the structure for a read-out chip after its separation from the dummy
  sensor. The {SLID} connections can be seen. The horizontal scale is
  \subref{fig:100mu} 100\,\mum\ and \subref{fig:200mu} 200\,\mum.
  \label{fig:rip_off_align}}
\end{figure}

 In \fig~\ref{fig:rip_off_align} photographs of the pulled off read-out chips
 are shown. In almost all cases the whole {SLID} stack is appreciable,
 indicating that the weakest point of the interconnection is at the
 electroplated layers, i.\,e.~layers that are similar in other technologies as
 for example bump bonding.
%
%
%
\section{Electrical Properties of the Pixel Modules}
 In the following the performance of the successfully built pixel modules from
 the chip-to-wafer prototype production are discussed based on results obtained
 before and after irradiation.
 These results comprise: leakage currents, tuning properties, charge collection
 measurements, and in addition hit efficiencies and cluster sizes determined in
 beam test measurements.
%
%
\subsection{IV Characteristics and Irradiation Programme}
\begin{figure}[h!tb]
\centering
\subfigure[]{
\includegraphics[width=0.47\textwidth]{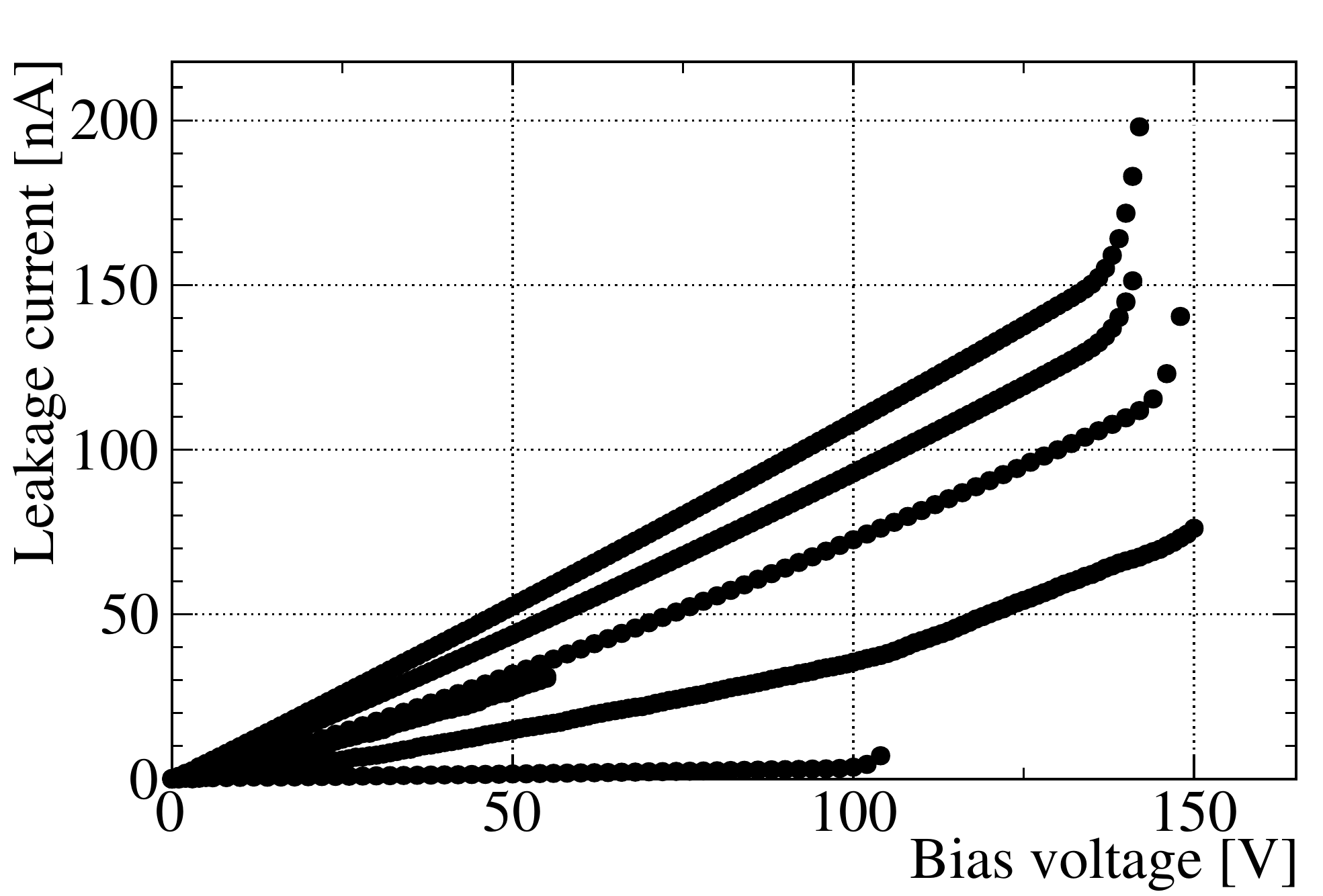}
\label{fig:FE-I2_IV}
}
\subfigure[]{
\includegraphics[width=0.47\textwidth]{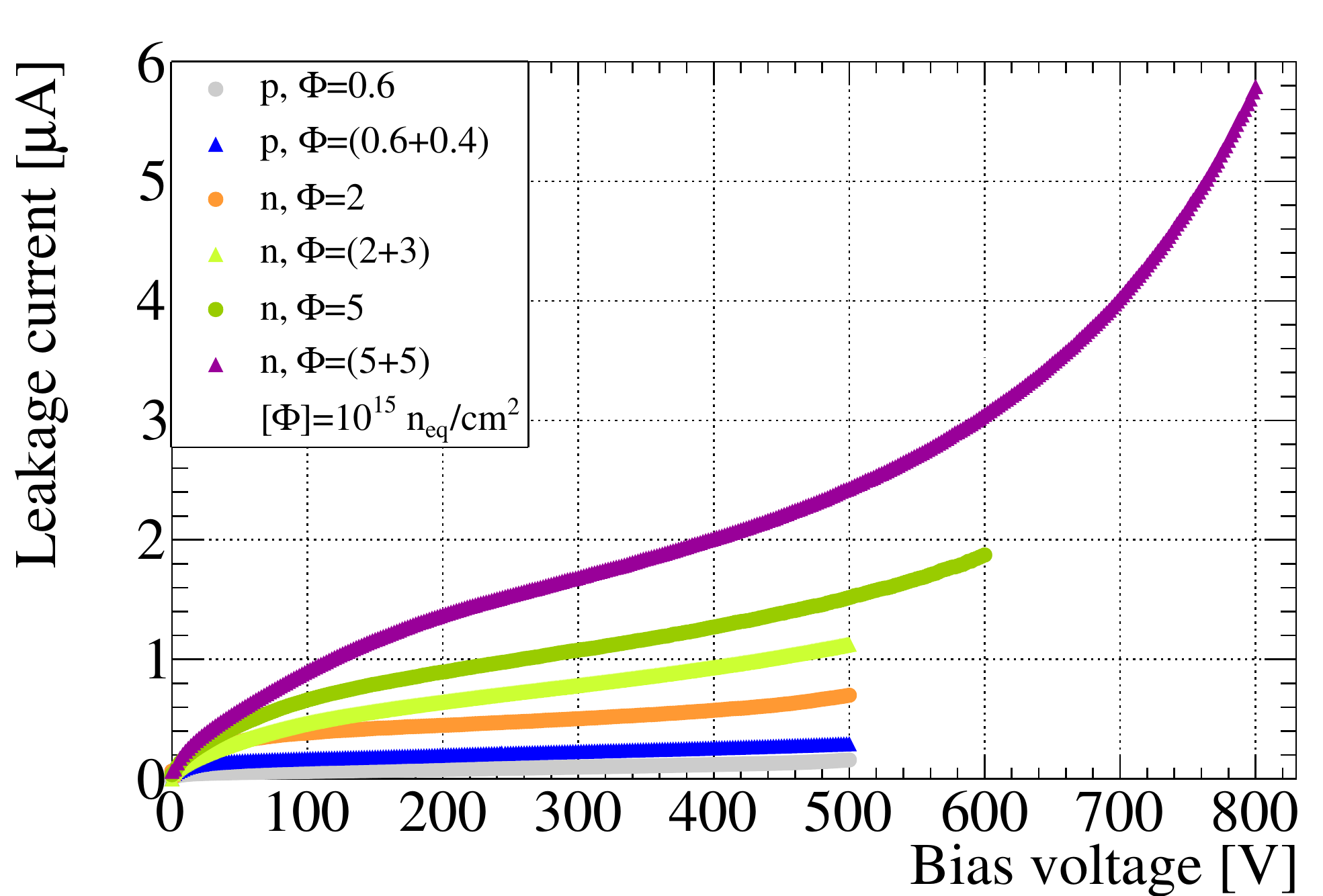}
\label{fig:IV_SLID_irrad}
}
\caption{IV characteristics \subref{fig:FE-I2_IV} before and
  \subref{fig:IV_SLID_irrad} after irradiation for the pixel modules. The
  curves for the two structures measured up to 55\,V before irradiation are
  indistinguishable. All measurements before (after) irradiation are taken at an
  environmental temperature of $20$\,\degC\ ($-50$\,\degC). The statistical
  uncertainties are smaller than the symbols.}
\end{figure}

 As basic functionality test, the IV characteristics of all seven modules are
 summarised in \fig~\ref{fig:FE-I2_IV}. All IV characteristics were taken with
 the read-out chip powered, but not configured, to ensure a defined ground
 potential and exclude temperature changes~\cite{NinPpaper}. At an
 over-depletion of about 10\,V, i.\,e.~at 30\,V, the leakage currents are below
 50\,nA and thus far below the operational limit of
 300\,$\mu$A~\cite{pixelelectronics}.

 The breakdown voltage lies for one module at 100\,V, for additional four
 modules at or above 140\,V. Additional two structures were measured only up to
 a bias voltage of 55\,V, and no breakdown was observed. For the structures
 measured up to the breakdown voltage, $V_{\mathrm{bd}}$, this corresponds to a
 good over-depletion ratio $V_{\mathrm{bd}}/V_{\mathrm{fd}}\geq3$.
%

 Subsequently, the modules were irradiated at the Karlsruhe Institut of
 Technology (KIT) with 25\,MeV protons~\cite{AldiPhD,Furgeri} and at the
 Jo\v{z}ef Stefan Institute (JSI) with reactor neutrons~\cite{Snoj2012483}. The
 full irradiation programme is summarised in
 \tab~\ref{tab:thinPix_Irradiations}. The range
 (0.6--10)$\cdot10^{15}$\,\neqcm\ was covered mainly with reactor neutron
 irradiation.
%
\begin{table}[ht]
\begin{center}
\begin{tabular}{|l|l|l|}
\hline
Fluence [$10^{15}$\,\neqcm]  & Irradiation site & Beam test\\ 
\hline
\hline
 0.6                           & KIT              & yes      \\ 
 0.6+0.4                       & KIT              &          \\ \hline
 2                             & JSI              & yes      \\ 
 2                             & JSI              &          \\ 
 2+3                           & JSI              &          \\ 
 5                             & JSI              & yes      \\ 
 5+5                           & JSI              &          \\ 
\hline
\end{tabular}
\caption[Overview of the received fluences for the irradiated modules and their
  respective irradiation sites.]{Overview of the received fluences for the
  irradiated modules and their respective irradiation sites. Assemblies tested
  in beam tests are indicated in the beam test column.
 \label{tab:thinPix_Irradiations}}
\end{center}
\end{table}

 In \fig~\ref{fig:IV_SLID_irrad} the leakage current as a function of the
 applied bias voltage is summarised for the irradiated assemblies. All
 measurements were taken at an ambient temperature of $-50$\,\degC\ to simulate
 as close as possible beam test environment temperatures where dry-ice cooling
 is employed. Again, the read-out chips were powered but not configured. The
 breakdown voltage of the irradiated sensors shifts to higher values and exceeds
 500\,V for all modules. Furthermore, the leakage currents are in agreement with
 expectations, showing increasing leakage currents with increasing
 fluences. Annealing effects are visible when comparing the module irradiated
 directly to a fluence of $5\cdot10^{15}$\,\neqcm\ with the module irradiated in
 two steps, since for irradiation at JSI an annealing time of about 1.5~days is
 unavoidable due to handling after each irradiation step. The latter module
 could not be investigated further, since the FE-I3 read-out chip failed after
 the second irradiation and remounting onto the test card. The leakage currents
 for all modules are found to be $\leq6\,\mu$A and thus again far below the
 operational limit of 300\,$\mu$A~\cite{pixelelectronics}.

 Assuming that all irradiated modules are fully depleted well below 450\,V, it
 was verified that the damage factors are about $6\cdot10^{-17}$\,A/cm,
 i.\,e.~in agreement with theory predictions for the different target fluences
 and received periods of annealing~\cite{MollPhD}.
%
%
\subsection{Module Tuning}
\label{sect:ThinTune}
 The module tuning and the charge collection measurements with radioactive
 sources were performed with the {ATLAS} {USBPix} read-out system~\cite{USBPix}.
 The expected most probable value (MPV) for the charge induced by
 $\beta$-electrons of the \Sr\ decay chain is about 4.9\,ke for the sensors with
 \dthin~\cite{BichselPDB,Bichsel1988}.  Therefore, the tuning is focussed on
 lowering the threshold as far as possible for each individual module. For the
 present FE-I3 read-out chip used in this R\&D programme thresholds down to
 3.2\,ke are generally achievable. For some single chip modules even lower
 thresholds down to (2.0--2.5)\,ke have been reached.
 This signal to threshold ratio is challenging for modules employing the present
 read-out chip.  However, results for the new ATLAS read-out chip FE-I4 show
 that it can be operated at thresholds as low as 1.6\,ke~\cite{MaltePSD9}, which
 is more than sufficient for the sensor thicknesses around 75\,\mum\ presented
 here.
 An additional complication for the prototype modules is imposed by the not
 connected pixel cells for some of the modules. This implies that in the tuning
 two very different states of the read-out chip in adjacent regions have to be
 accommodated.
%
\begin{figure}[h!tb]
\centering
\subfigure[]{
\includegraphics[width=0.47\textwidth]{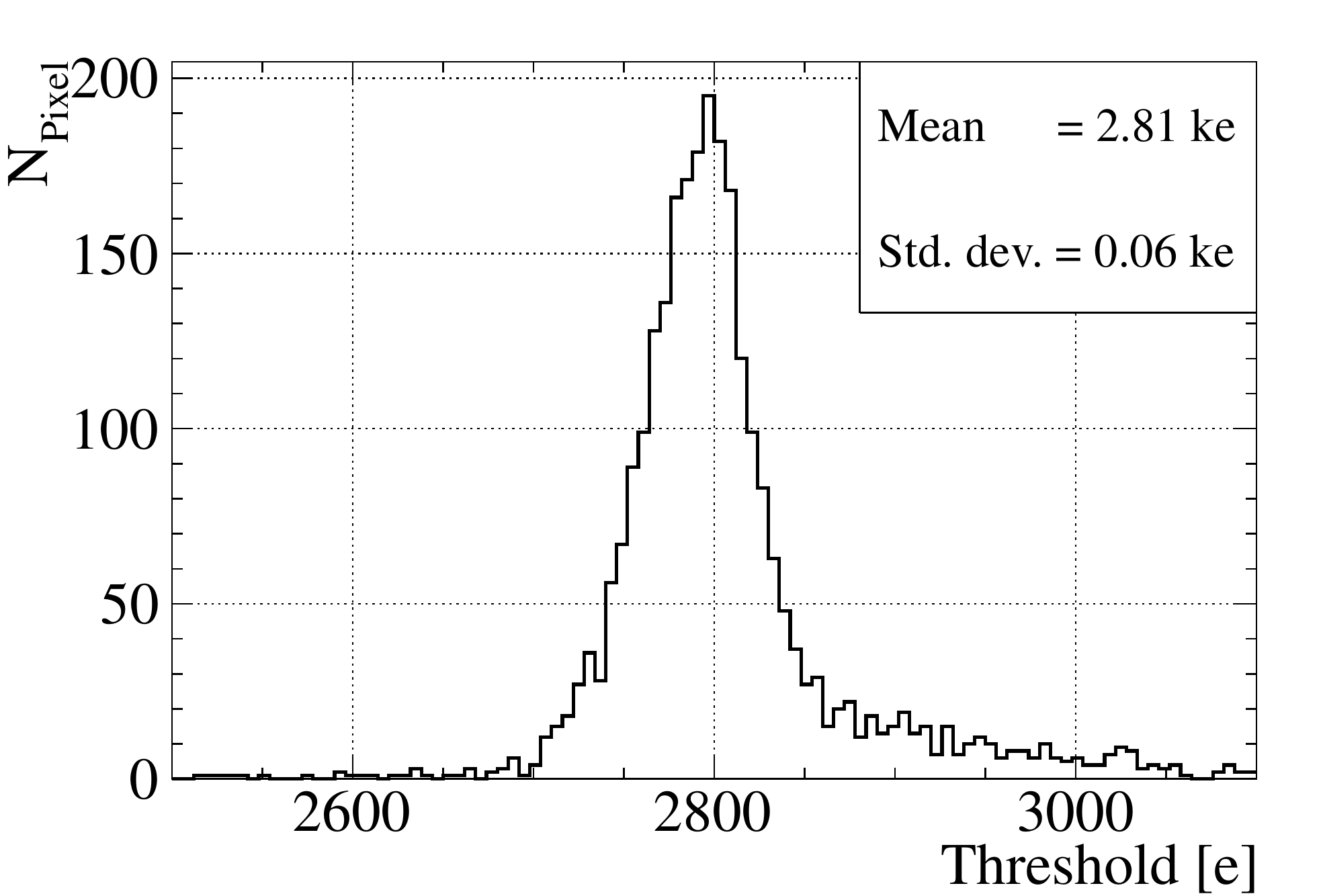}
\label{fig:SLID10_THR2800_Thr}
}
\subfigure[]{
\includegraphics[width=0.47\textwidth]{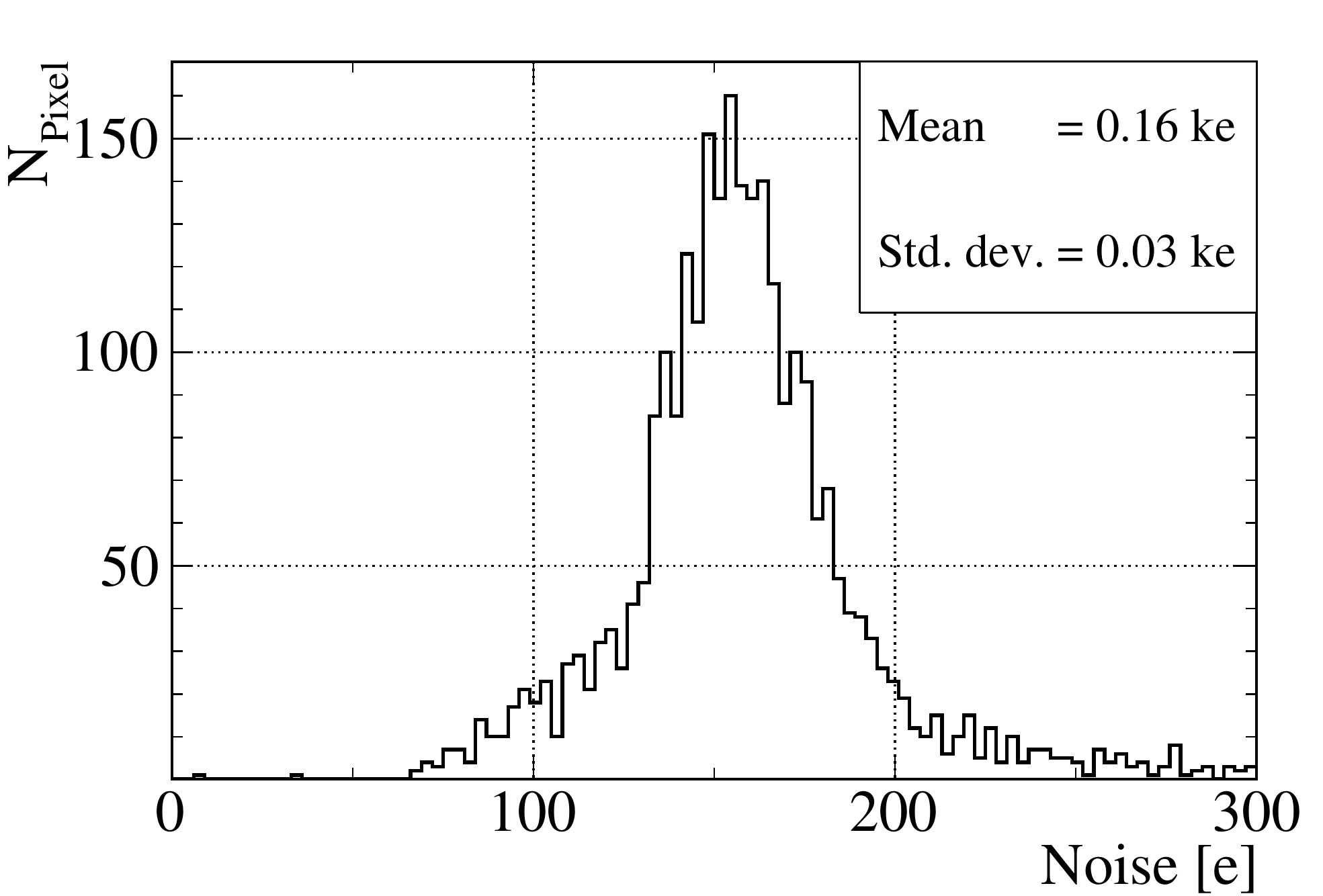}
\label{fig:SLID10_Thr2800_Noise}
}
\caption{Typical \subref{fig:SLID10_THR2800_Thr} threshold and
  \subref{fig:SLID10_Thr2800_Noise} noise distribution for a prototype pixel
  module.}
\label{fig:SLID10_Thr2800}
\end{figure}

 The threshold and noise distributions for a typical tuning are shown in
 \fig~\ref{fig:SLID10_Thr2800}.
 The target threshold of 2.8\,ke was reached for about 90$\%$ of the pixel cells
 with a standard deviation of 0.06\,ke. The corresponding noise is 0.16\,ke with
 a standard deviation of 0.03\,ke over the module and thus not significantly
 different from the noise found for other n-in-n and n-in-p modules with
 thicknesses in the range (250-285)~\mum~\cite{pixelelectronics,NinPpaper}.
%
\begin{figure}[h!tb]
\centering \subfigure[]{
  \includegraphics[width=0.47\textwidth]{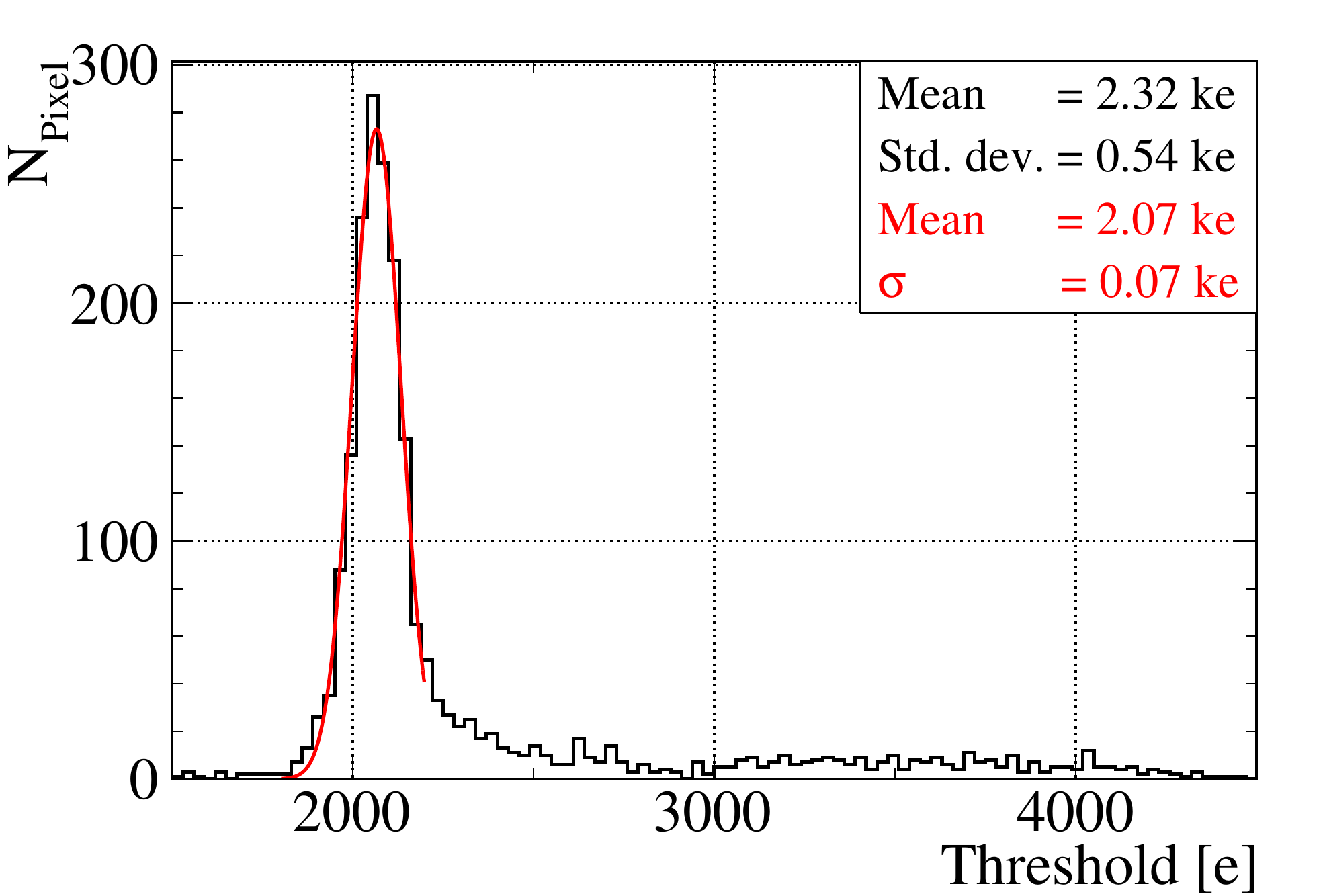}
\label{fig:SLID3_1e16_Thr2100_Thr}
} \subfigure[]{
  \includegraphics[width=0.47\textwidth]{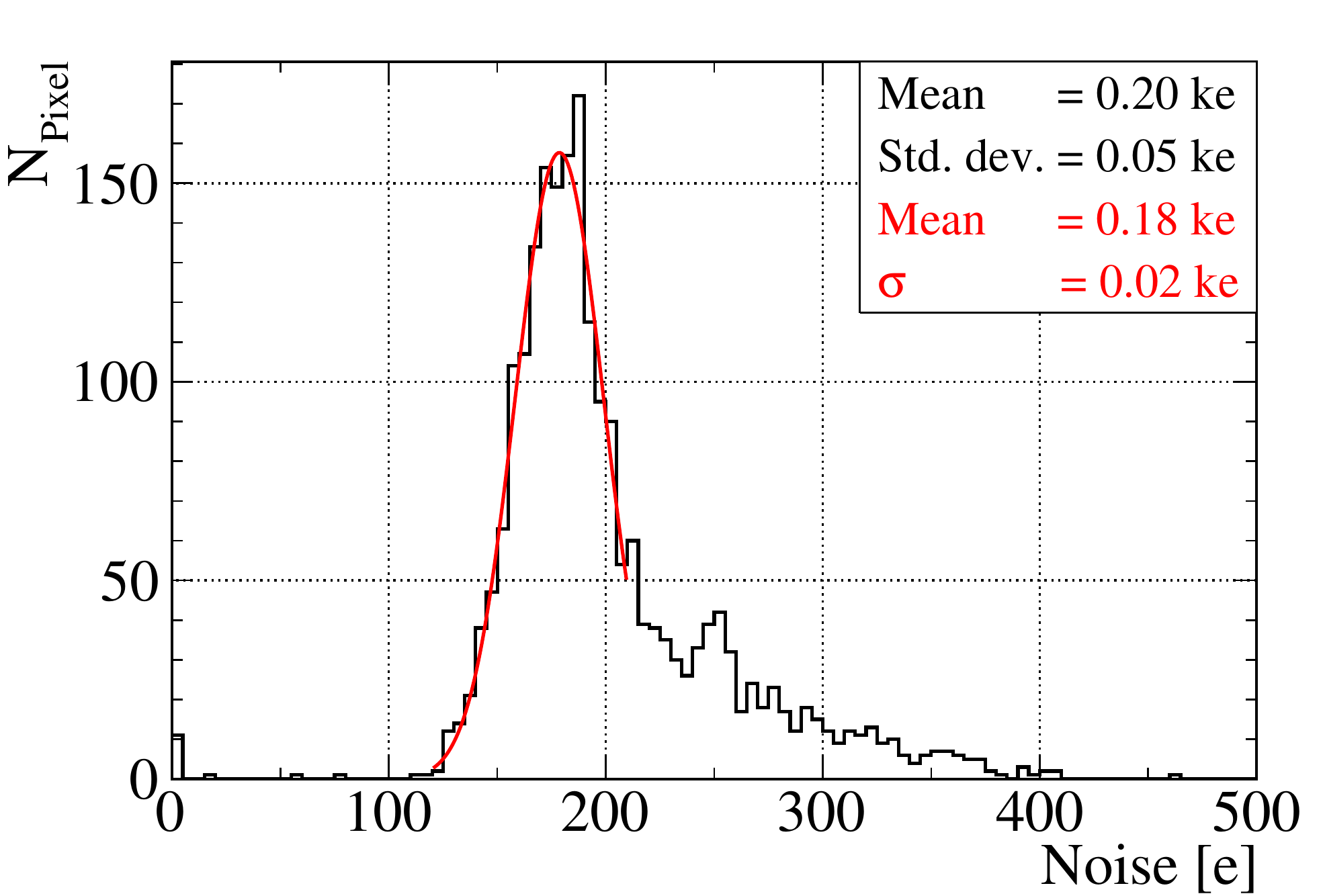}
\label{fig:SLID3_1e16_Thr2100_Noise}
} \subfigure[]{
  \includegraphics[width=0.47\textwidth]{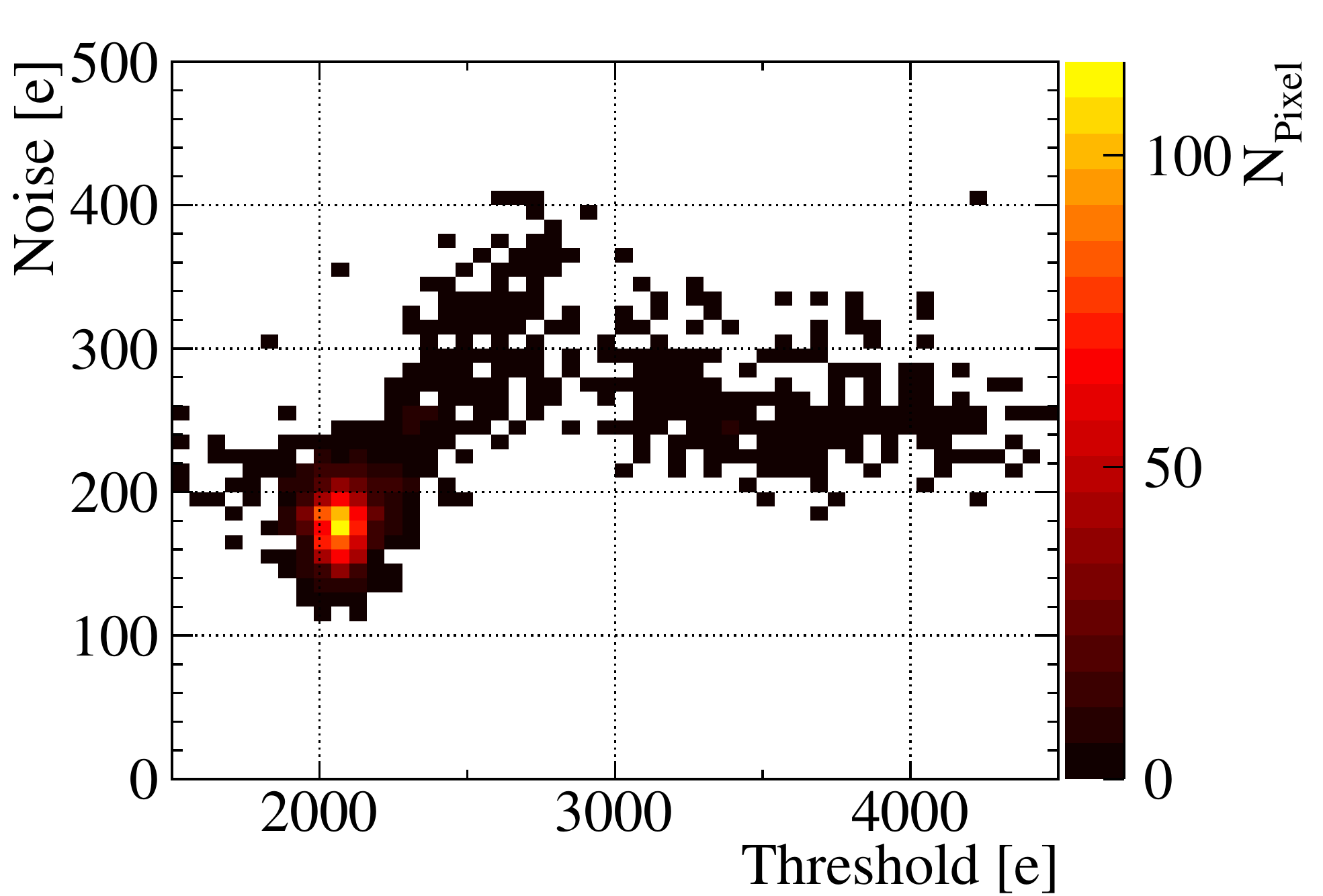}
\label{fig:SLID3_1e16_Thr2100_NoisevsThr_Correlation_Th2}
}
\caption[Threshold and noise distribution for the module irradiated to
  $10^{16}$\,\neqcm.]{\subref{fig:SLID3_1e16_Thr2100_Thr} Threshold and
  \subref{fig:SLID3_1e16_Thr2100_Noise} noise distribution for the module
  irradiated to $10^{16}$\,\neqcm. In
  \subref{fig:SLID3_1e16_Thr2100_NoisevsThr_Correlation_Th2} the pixel-by-pixel
  correlation of threshold and noise values is given.}
\label{fig:SLID3_1e16_Thr2100}
\end{figure}

 In \fig~\ref{fig:SLID3_1e16_Thr2100} the results of the tuning with the lowest
 achieved threshold and the corresponding noise among all modules before and
 after irradiation is depicted. It was achieved for the module irradiated to a
 fluence of $10^{16}$\,\neqcm.  The mean threshold, shown in
 \fig~\ref{fig:SLID3_1e16_Thr2100_Thr}, was tuned as low as $2.32$\,ke with a
 standard deviation of $0.54$\,ke across the module. The corresponding noise,
 shown in \fig~\ref{fig:SLID3_1e16_Thr2100_Noise}, is $0.20$\,ke with a standard
 deviation of $0.05$\,ke across the module. The long tail of the distributions
 is mainly caused by pixel cells which could not be tuned to such low
 thresholds.  The pixel-by-pixel correlation of threshold and noise, shown in
 \fig~\ref{fig:SLID3_1e16_Thr2100_NoisevsThr_Correlation_Th2}, demonstrates that
 the outliers in both distributions coincide. Since this is a known issue of the
 FE-I3 read-out chip, which is not planned to be used for future ATLAS upgrades,
 these outlier pixel cells are disregarded in the following. 
 Fitting a Gaussian to the core of the distributions, for the tuning shown in
 Figures~\ref{fig:SLID3_1e16_Thr2100_Thr}
 and~\ref{fig:SLID3_1e16_Thr2100_Noise}, the threshold lies at
 $(2.07\pm0.07)$\,ke and the corresponding noise is $(0.18\pm0.02)$\,ke.
%
\begin{figure}[tbh]
\centering
\includegraphics[width=0.47\textwidth]{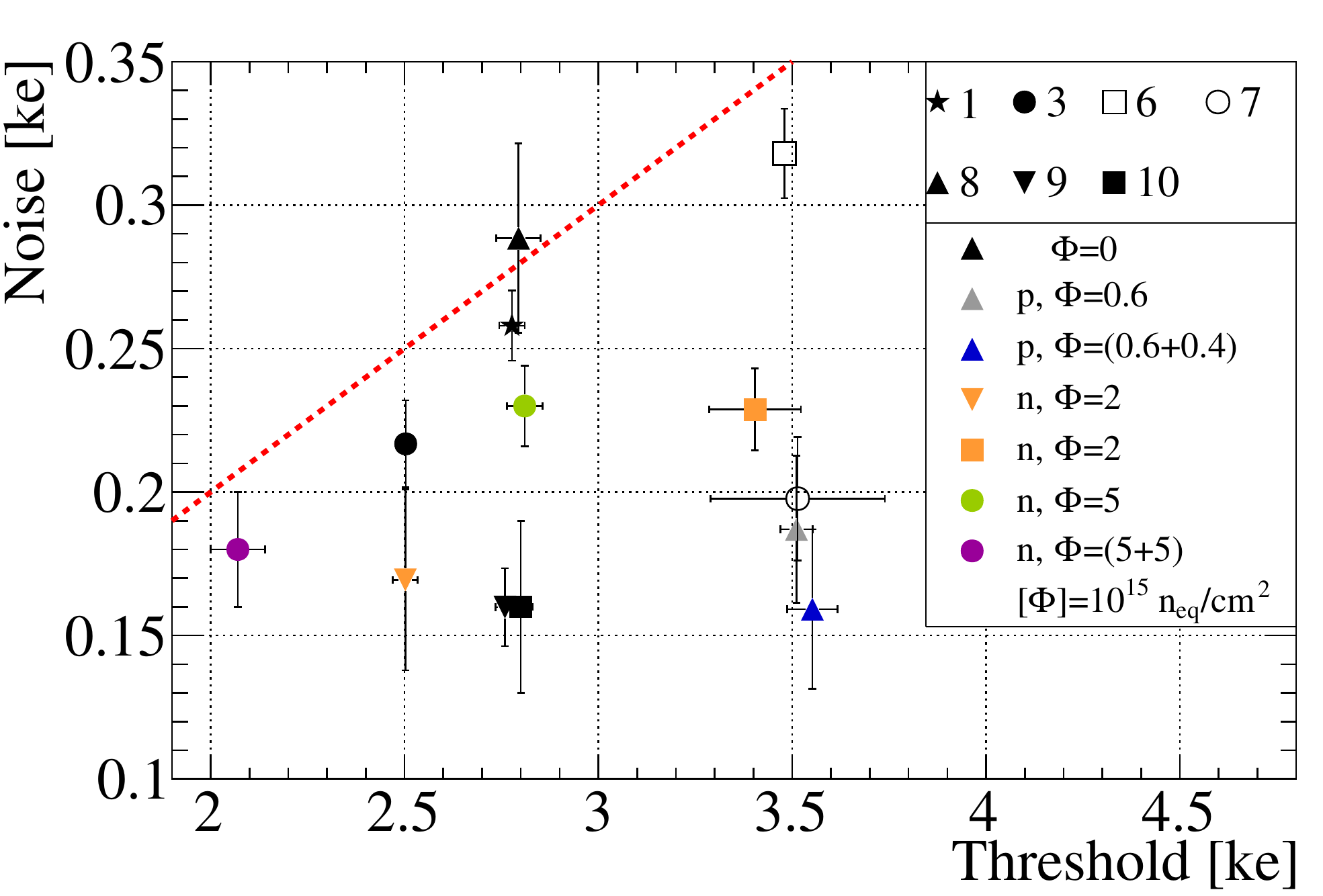}
\caption[Best achieved mean thresholds and their respective noise values for the
  modules before and after irradiation.]{Best achieved mean thresholds and their
  respective noise values for the modules before and after irradiation. The
  symbol style denotes the module and the colour the received
  fluence. Irradiation with protons (neutrons) are indicated by p (n).  For
  better visibility, the data point for module 9 before irradiation has been
  slightly displaced horizontally by -30\,electrons. The uncertainties indicate
  the standard deviations of the respective distributions. Before irradiation
  the environment temperature is kept at 20\,\degC, afterwards at
  $-50$\,\degC. The red dotted line indicates a threshold to noise ratio of
  ten. (For interpretation of the references to colour in this figure caption,
  the reader is referred to the web version of this paper.)}
\label{fig:FE-I2_Thr_Noise}
\end{figure}

 An overview of the threshold tuning and corresponding noise values of all
 modules before and after irradiation is given in
 \fig~\ref{fig:FE-I2_Thr_Noise}, where the lowest achieved thresholds and their
 corresponding noise values are given for each module. The uncertainties shown
 correspond to the standard deviation of threshold and noise, respectively. The
 average noise observed for all assemblies is $(0.21\pm0.01)$\,ke. The slightly
 increased value with respect to currently used modules is due to the lower
 threshold target values and the influence of the not connected pixel cells. The
 effect of the not connected pixel cells is especially pronounced in the
 assemblies with the highest number of not connected cells, number 6 (open
 squares) and 7 (open circles). Nonetheless, an excellent threshold to noise
 ratio exceeding ten (red dotted line) in all but one case is achieved for
 assemblies before as well as after irradiation.
%
%
\subsection{Charge Collection}
\label{sect:ThinPixQ}
 Thin sensors show a higher CCE after irradiation, since the full depletion
 voltages are reduced, and higher electric fields are achieved when applying the
 same bias voltage. 
 To investigate the charge collection, measurements using either photons from an
 \Am\ source, or $\beta$-electrons from a \Sr\ source, were conducted.
 While for photons the internal trigger logic was used, for $\beta$-electrons an
 external trigger was employed.
 Within uncertainties no significant difference in charge collection was found
 between the modules.
%
%
\subsubsection{Radioactive Source Measurements}
\begin{figure}[tbh!]
\centering
\includegraphics[width=0.47\textwidth]{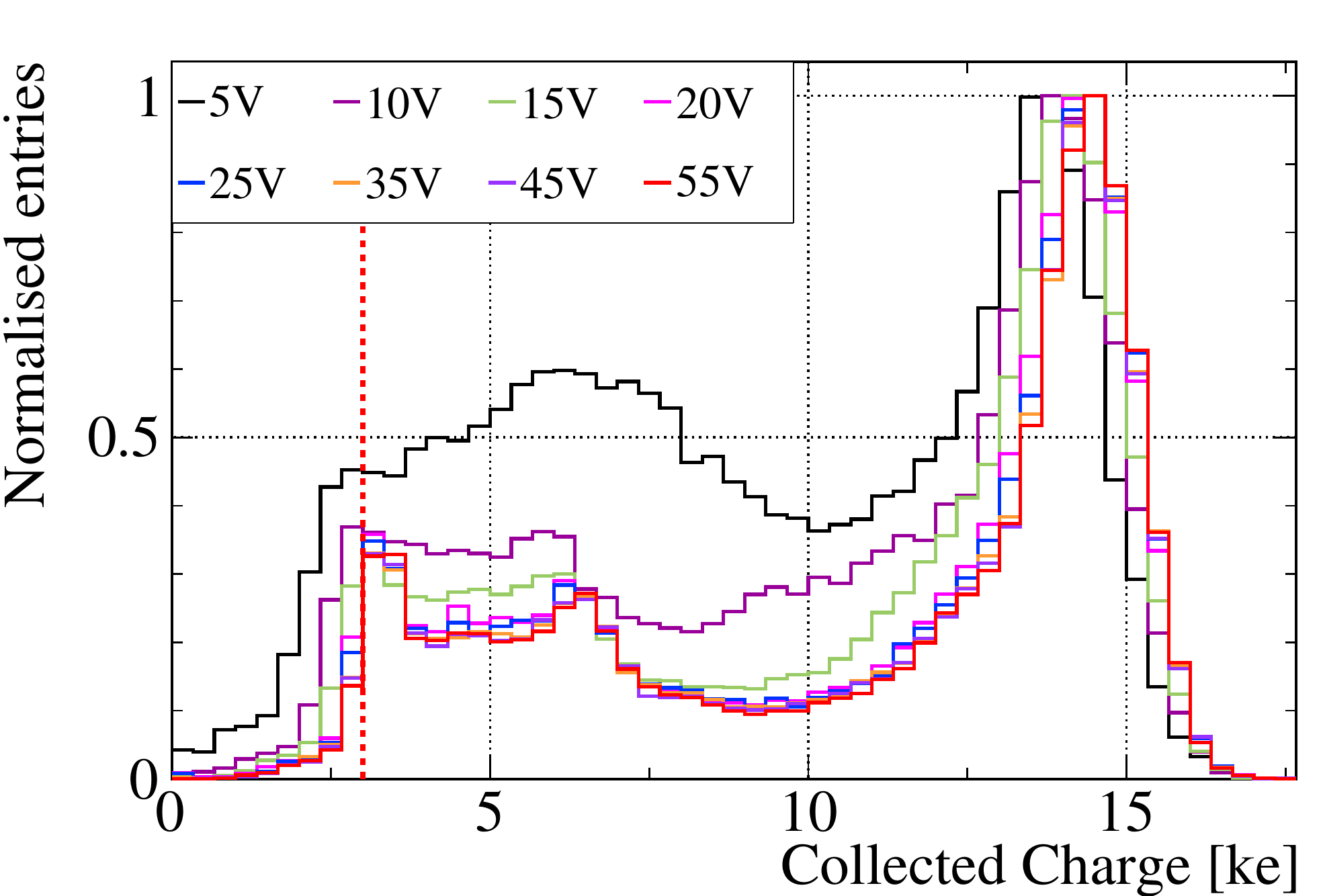}
\caption[Evolution of an \Am\ source energy spectrum with the applied bias
  voltage for a module.]{Evolution of an \Am\ source energy spectrum with the
  applied bias voltage. The threshold of 3.0\,ke is indicated by the red dotted
  line. (For interpretation of the references to colour in this figure caption,
  the reader is referred to the web version of this paper.)
  \label{fig:SLID3Am241}}
\end{figure}

 In \fig~\ref{fig:SLID3Am241} the \Am\ photon spectra obtained with a module
 biased at different bias voltages between 5\,V and 55\,V are depicted. Each
 histogram is normalised to its bin with the highest content. For a high
 resolution reference spectrum taken with a high purity Germanium detector
 please refer to~\cite{Am241_pub,Am241_spectrum}.  At 55\,V the prominent
 59\,keV $\gamma$-line is measured at
 ($14.4\pm0.5{\mathrm{(fit)}}\pm1.1{\mathrm{(syst.)}}$)\,ke (Gaussian fit not
 shown), which is in good agreement with the expected peak position of 16.4\,ke,
 when taking into account the calibration bias of the FE-I3 read-out
 chip~\cite{BONN-IR-2008-04}.
 The first uncertainty denotes the one from the fit, and the second the
 systematic uncertainty stemming from the charge calibration of the read-out
 chip. The second prominent line in the spectrum at 26\,keV is expected at
 7.2\,ke. However, due to the charge resolution it merges with the lines below,
 such that only the upper edge is appreciable, which lies between 6\,ke and
 7.5\,ke.
%

 At lower bias voltages a fraction of the sensor volume does not contribute to
 the charge collection and thus the full amount of charge is only collected for
 events where the photo-electric process occurred in an already depleted region.
 For those events where it instead occurs in the not yet depleted part, only the
 fraction of the charges diffusing into the depleted volume can be
 measured. This leads to a broadening of the peaks and to a less defined
 spectrum. Due to the small thickness of the sensor, only the measurements below
 15\,V are significantly affected.
%
\begin{figure}[h!tb]
\centering
\subfigure[]{
\includegraphics[width=0.47\textwidth]{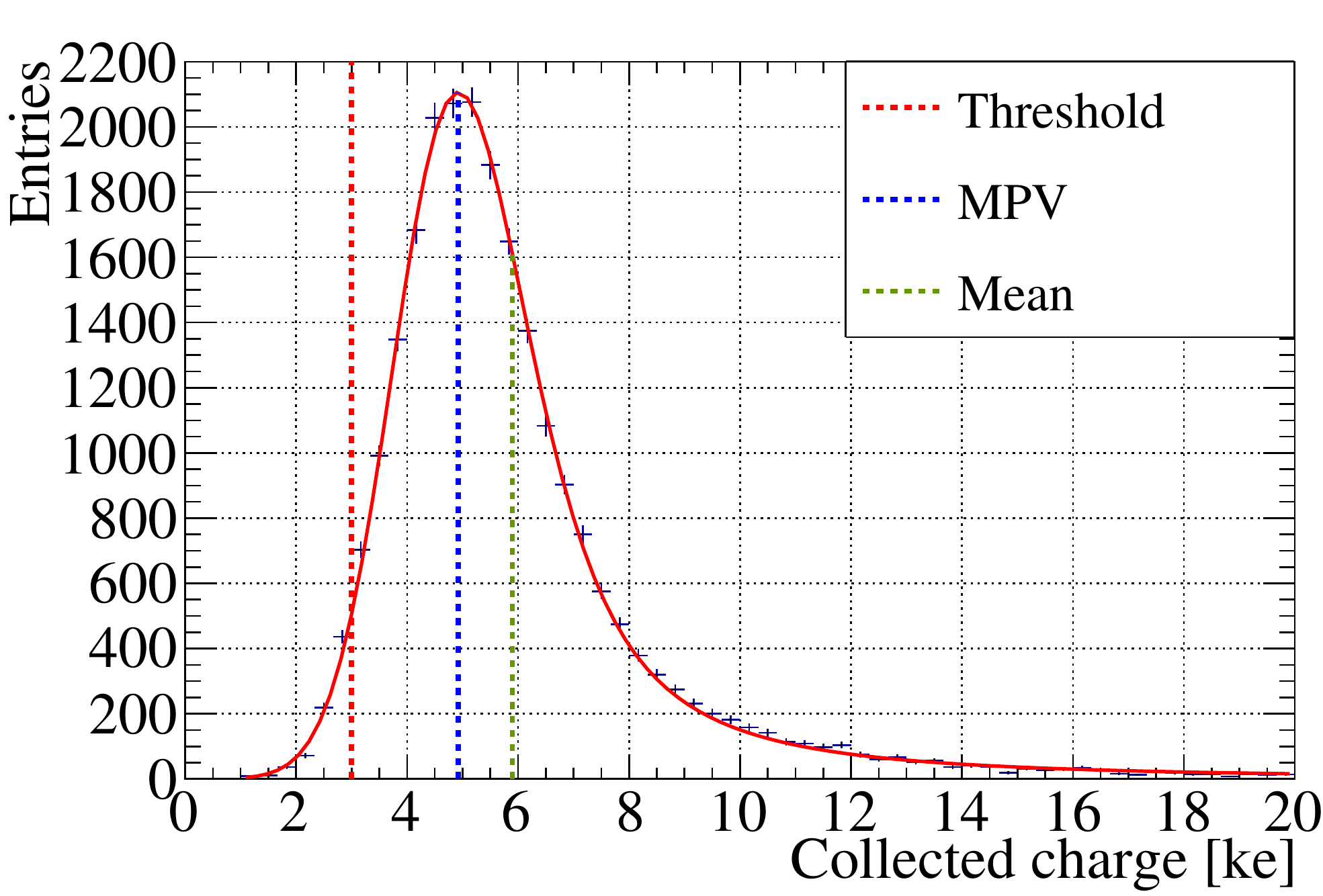}
\label{fig:SLID3Landau}
}
\subfigure[]{
\includegraphics[width=0.47\textwidth]{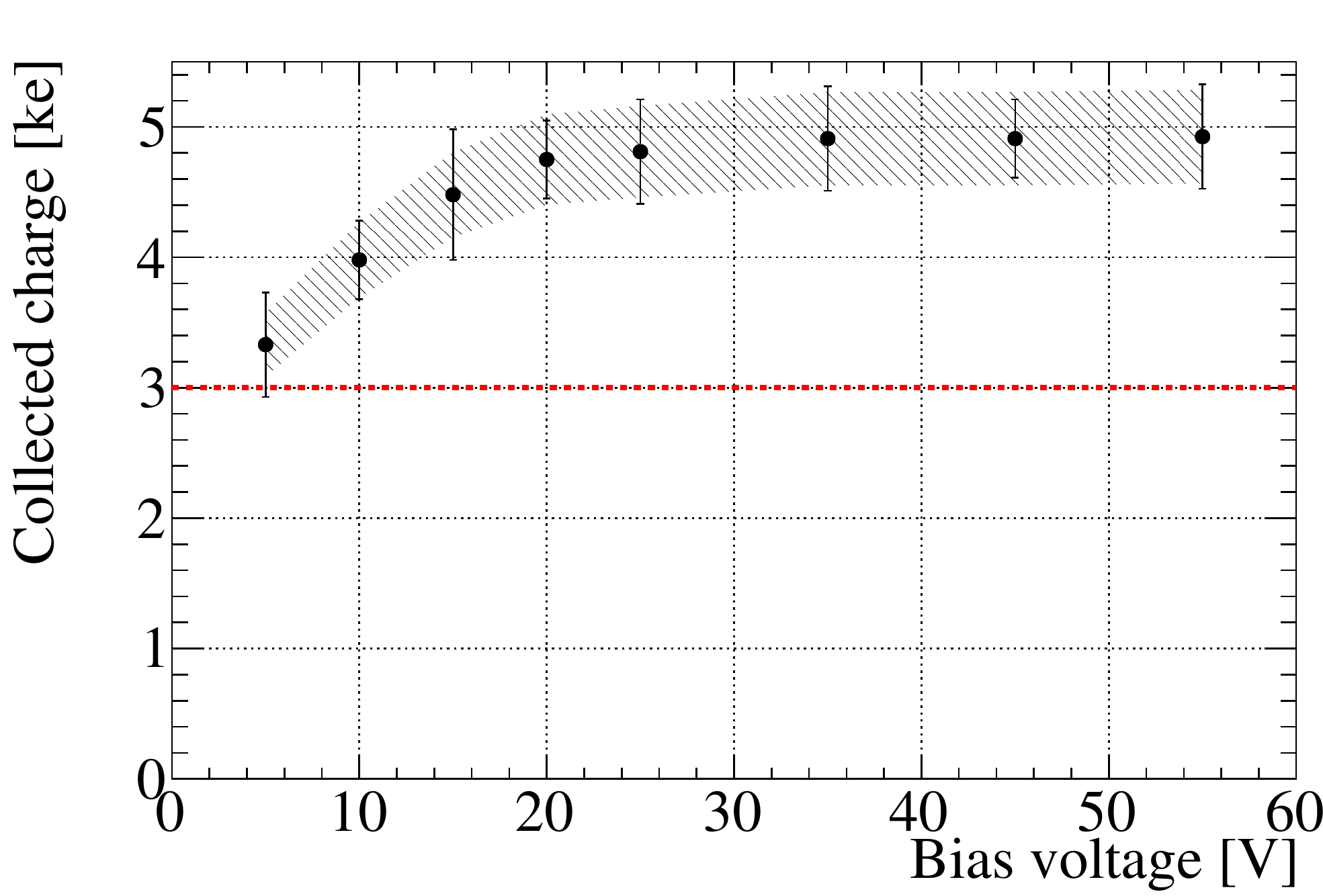}
\label{fig:SLID3_QvsV}
}
\caption[]{\subref{fig:SLID3Landau} Distribution of collected charges induced by a
  \Sr\ source for a module biased at 55\,V. The data are shown with their
  statistical uncertainties. The fit function, a Landau distribution convoluted
  with a Gaussian, is shown in red. The three vertical lines denote the
  threshold~(red), the MPV~(blue), and finally, the mean~(green) of the
  distribution within the range 1--20\,ke.

  The resulting MPV of the collected charges as a function of the bias voltage
  is shown in \subref{fig:SLID3_QvsV}. The uncertainty bars account for the
  fitting uncertainty and the band for the fully correlated systematic
  uncertainty. The dotted red line indicates the threshold. (For interpretation
  of the references to colour in this figure caption, the reader is referred to
  the web version of this paper.)}
\label{fig:Charge_SourceScan}
\end{figure}

 A charge distribution of a \Sr\ measurement of a module operated at a bias
 voltage of 55\,V is shown in \fig~\ref{fig:SLID3Landau}. The threshold in this
 measurement was tuned to 3.0\,ke and is indicated by the red dotted
 line. Entries below threshold occur because the threshold corresponds to the
 50$\%$ efficiency point. In addition indicated are the MPV (blue line) and the
 mean value (green line) of the collected charge.
 The measurement is well described by a convolution of a Landau distribution
 with a Gaussian. The fit, based on the statistical uncertainties of the data,
 was performed in the range 1--20\,ke.
 The evolution of the resulting MPV of the collected charge as a function of the
 bias voltage is summarised in \fig~\ref{fig:SLID3_QvsV}.
 The uncertainty, shown as a band, is fully correlated from point to point and
 caused by the calibration uncertainty. Since the MPV of the collected charge is
 close to the threshold, the uncertainties arising from the fit are increased
 due to the deformation of the Landau distribution. They are indicated by the
 uncertainty bars. Full charge collection is reached at a bias voltage of
 $(21\pm0.7)$\,V as determined by the intersection of two linear functions
 describing the different parts of the charge collection measurement.
%
\begin{figure}[tbh!]
\centering
\subfigure[]{
\includegraphics[width=0.47\textwidth]{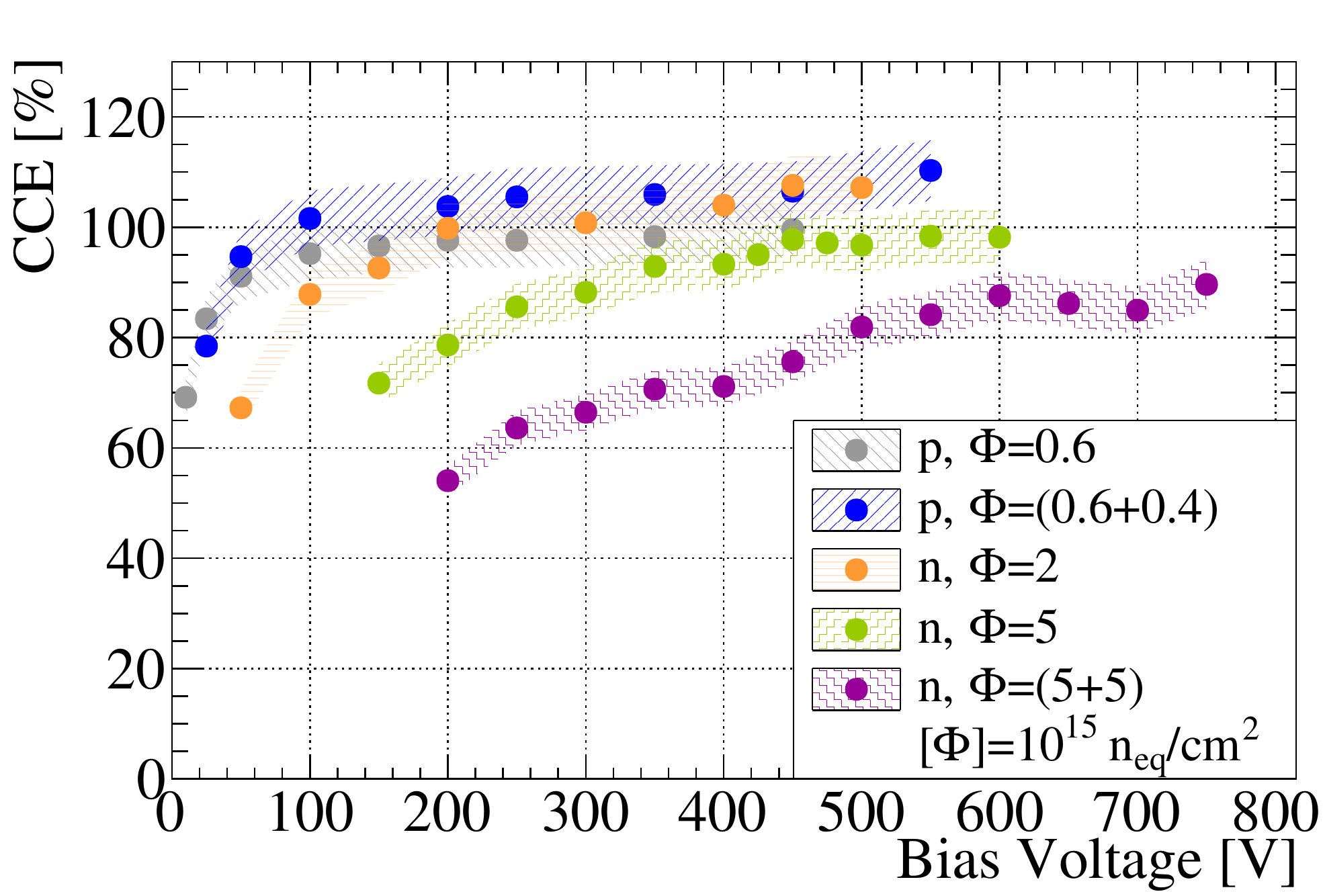}
\label{fig:SLID_irrad_CCE}
}
\subfigure[]{
\includegraphics[width=0.47\textwidth]{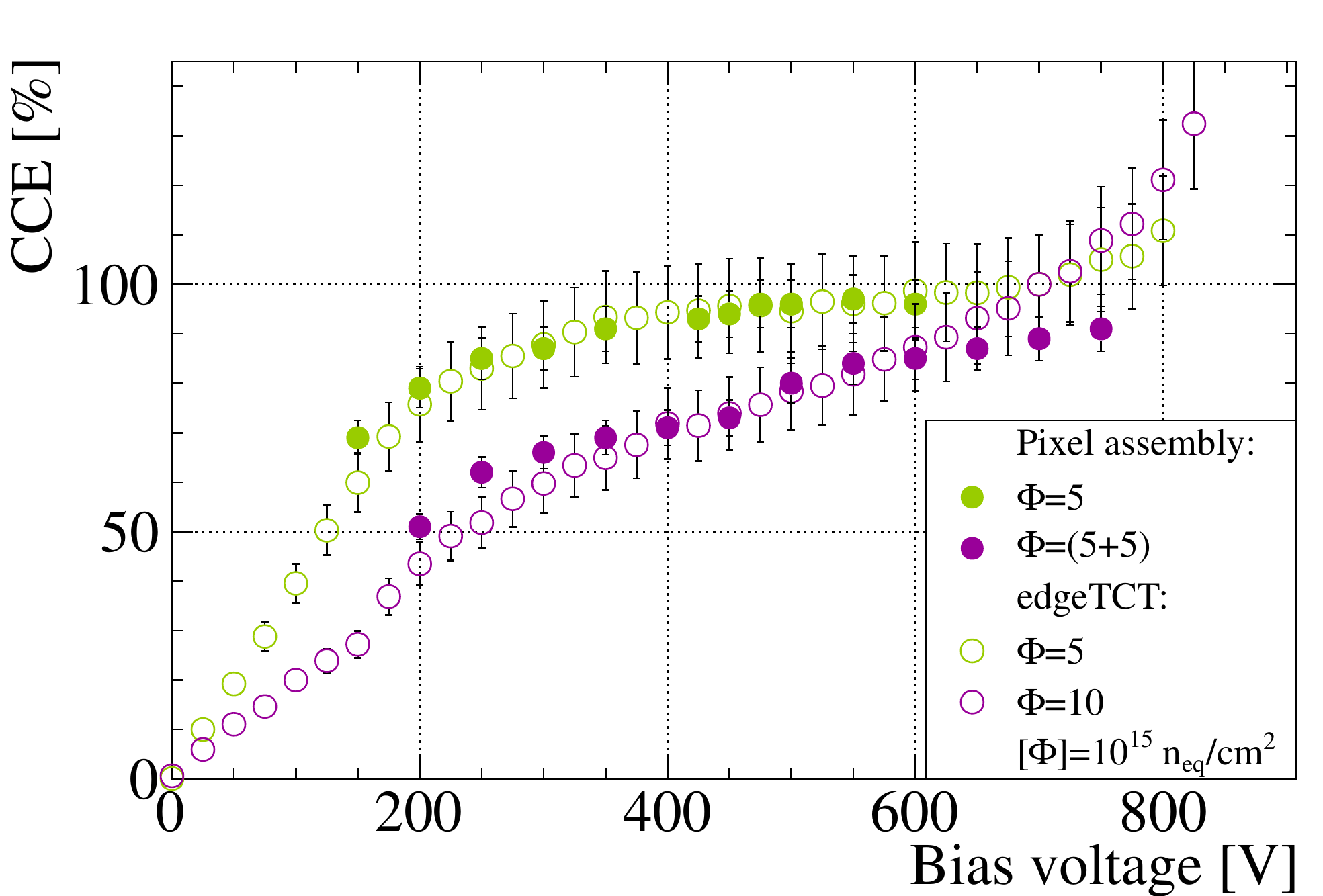}
\label{fig:edgevspix}
}
\caption{\subref{fig:SLID_irrad_CCE} CCE with respect to the maximum charge
  collected by the respective module before irradiation as a function of the
  applied bias voltage for irradiated modules. Proton (neutron) irradiated
  samples are denoted with p (n) in the legend. The uncertainty band accounts
  for the overall time-over-threshold to charge
  calibration. \subref{fig:edgevspix} Comparison of the MPV of the collected
  charges obtained with pixel modules (full symbols) to infra-red laser
  measurements on strip sensors from the same production~\cite{WeigellPhD} (open
  symbols). For a better visibility the fully correlated uncertainty bands are
  drawn as simple bars. (For interpretation of the references to colour in this
  figure caption, the reader is referred to the web version of this paper.)}
\label{fig:SLIDirradcharge}
\end{figure}
%
 This agrees well with the infra-red laser measurements on strips from the same
 production. For the module shown the charge saturates at
 $(4.6\pm0.4(\mathrm{fit})\pm0.3(\mathrm{syst.}))$\,ke and thus is in good
 agreement with the expectations for a sensor with \dthin.

 Aiming for usage at the expected HL-LHC environment, a high CCE at high
 irradiation levels is of utmost importance. The measured values of this
 parameter are summarised for all irradiated modules as a function of the
 applied bias voltage and for different received fluences (colour) in
 \fig~\ref{fig:SLID_irrad_CCE}.
 Since the uncertainties stemming from the charge calibration before and after
 irradiation are highly correlated they almost completely cancel, when
 investigating the ratio. Still, as a conservative estimate a 5$\%$ uncertainty
 is assigned to the ratio. As expected from the strip
 measurements~\cite{WeigellPhD}, within the assessable voltage range, a
 saturation is found up to the highest fluences. The onset of the saturation
 increases with fluence, but lies at comparably low voltages for all fluences,
 i.\,e.~below 500\,V. These low bias voltages, in combination with the fact that
 all modules saturate within uncertainties to a CCE of 100$\%$ up to a received
 fluence of $5\cdot10^{15}$\,\neqcm and to 90$\%$ at a received fluence of
 $10^{16}$\,\neqcm, allows to operate them in a restricted bias voltage range
 over the entire life-time of an experiment. This leads to looser requirements
 on the read-out electronics.

 For comparison in \fig~\ref{fig:edgevspix} the results obtained from infra-red
 laser measurements on strip sensors from the same production are depicted
 together with the results obtained with the pixel modules for the two highest
 received fluences~\cite{WeigellPhD}. For this figure the infra-red laser
 measurements were renormalised globally to achieve comparable scales. For the
 measurement at $5\cdot10^{15}$\,\neqcm\ an excellent agreement is found over
 the entire range. At the higher fluence slight deviations at low and high
 applied bias voltages are observed, which are most likely caused by the
 different annealing history of the structures, given the two step irradiation
 procedure for the pixel module. However, considering the use of a single
 scaling factor over the entire range, a good agreement is achieved.
%
%
\subsection{Cluster Size}
\label{sect:ThinPixSpatRes}
 The cluster size and hit efficiency were determined with test beam data
 obtained with 120\,GeV pions at the CERN SPS.
 The position within a given pixel assembly under test where the particle
 traverses the assembly is determined from external information provided by the
 EUDET beam telescope~\cite{EUDET}.
 Analysing the signals from the pixels around this position the cluster size and
 the hit efficiency can be determined.
%
\begin{figure}[t]
\centering
\includegraphics[width=0.47\textwidth]{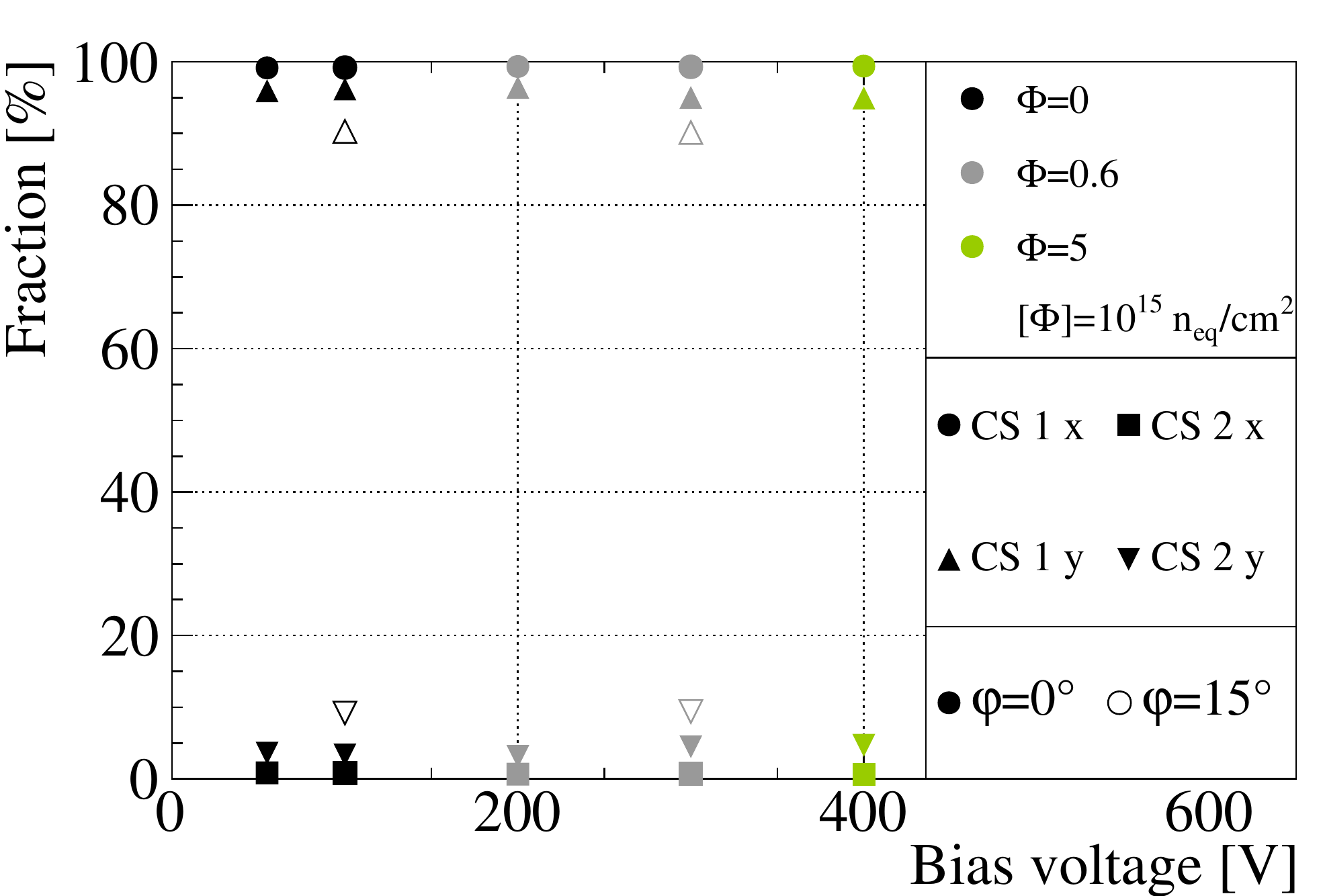}
\caption{Summary on the cluster size fractions as a function of the bias voltage
  before and after irradiation. The uncertainties are calculated according
  to~\protect\cite{Paterno} and are smaller than the symbol sizes. The colours
  represent the received fluences and the marker type the cluster size as well
  as its spatial coordinate. Filled (open) markers stand for measurement at
  perpendicular ($\varphi=15^{\circ}$) incidence. (For interpretation of the
  references to colour in this figure caption, the reader is referred to the web
  version of this paper.)}
\label{fig:FE-IXCS_overview}
\end{figure}

 For thinner sensors the spatial resolution is expected to differ from the one
 observed for thick sensors given the different cluster size
 abundances. However, when comparing the resolution on events with a specific
 cluster size between different thicknesses no difference is expected. In any
 case, lower cluster sizes lead to a reduced occupancy.
%
\begin{figure*}[ht!]
\centering
\subfigure[]{
\includegraphics[width=0.98\textwidth]{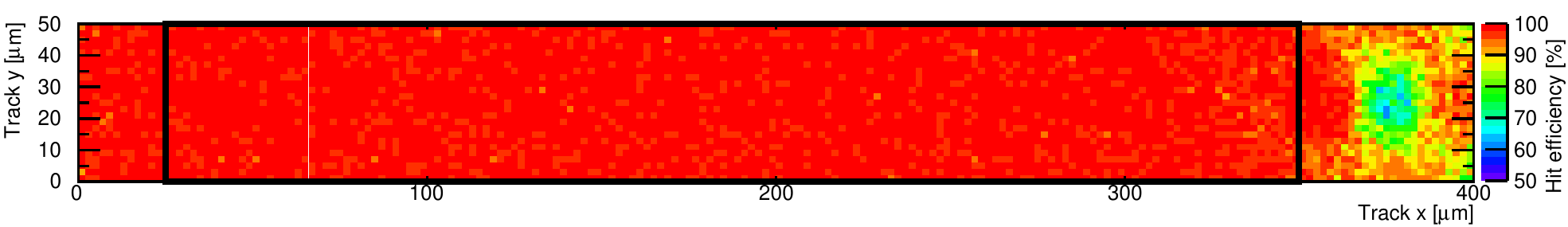}
\label{fig:SLID9eff100V0deg}
}
\subfigure[]{
\includegraphics[width=0.92\textwidth]{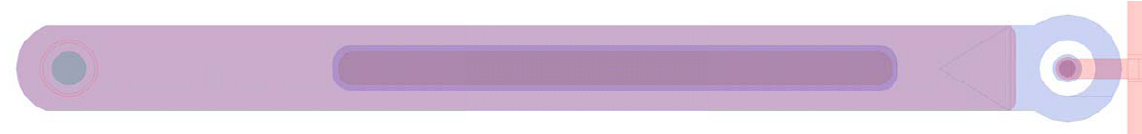}
\label{fig:pixel-layer}
}
\caption{\subref{fig:SLID9eff100V0deg} Map of the mean hit efficiency as a
  function of the impact point predicted by the beam telescope at a bias voltage
  of 100\,V. For reference in \subref{fig:pixel-layer} the design of a single
  pixel cell is given. The implantation extends over the entire structure shown,
  and has a ring shaped opening at the punch through bias dot displayed on the
  right side. The metal layer, covering most of the implant, is shown as a large
  rectangle with rounded corners on the left side. The T-shaped structure at the
  far right end comprises the metal lines, connecting the bias dot to the bias
  ring. The opening in the nitride and oxide layers is displayed as the
  rectangle in the centre of the pixel. The small circle at the left end of the
  pixel is the opening in the passivation, where the pixel will be connected
  with bump bonding.}
\label{fig:pixeleff}
\end{figure*}

 The low absolute collected charge to threshold ratio is reflected in the
 smaller abundances of higher multiplicity clusters. In
 \fig~\ref{fig:FE-IXCS_overview} a summary of the cluster size as a function of
 the bias voltage for different received fluences (colour), spatial coordinates
 (symbol style), and particle incidence angles $\varphi$ (closed/open symbols)
 is given. The uncertainties are calculated according to~\cite{Paterno} and are
 smaller than the symbol sizes. In the direction of the short pixel pitch $y$
 only about 5$\%$ of two-hit clusters are observed for perpendicular
 incidence. If the modules are tilted by $\varphi=15^{\circ}$, as it is foreseen
 for the IBL, about 10$\%$ of the clusters in $y$ are composed of two hits. As
 expected from the CCE measurements, no difference is found before and after
 irradiation, provided that the applied bias voltage is around or above the
 value corresponding to the charge saturation.
%
%
\subsection{Hit Efficiency}
\label{sect:ThinPixEff}
 Besides the resolution, the tracking efficiency of the pixel detector is the
 key figure of merit. For a high tracking efficiency, a high hit efficiency of
 the pixel assemblies is mandatory. The latter is mainly driven by the ratio
 between collected charge and threshold. Consequently, for thinner sensors the
 lowest possible threshold is desirable as discussed before. This criterion is
 especially challenging since the difference between the mean threshold and the
 {MPV} of the collected charge is so small that a part of the distribution lies
 below threshold, as shown for example in \fig~\ref{fig:SLID3Landau}. Since the
 threshold corresponds to an efficiency of 50$\%$ for the electronic circuit of
 the pixel cell this considerably diminishes the overall hit efficiency. In
 \fig~\ref{fig:SLID9eff100V0deg} the mean hit efficiency as a function of the
 impact point predicted by the beam telescope is depicted for a bias voltage of
 100\,V. For this measurement, the thresholds were tuned to 2800\,e.

 The impact of the comparably high threshold is most pronounced in the area of
 the punch-through bias structure, and in the corner regions, where it leads to
 a loss of hit efficiency due to the sharing among several pixels.
 Anyhow, both effects are most pronounced for perpendicular impinging particles,
 occurring only for the very central part of a high energy physics
 experiment. Therefore, the quoted hit efficiency has to be understood as a
 lower bound. The overall hit efficiency is found to be $(98.1\pm0.3)\%$. If
 just the central region, indicated by the box in
 \fig~\ref{fig:SLID9eff100V0deg}, is considered, the hit efficiency rises to
 $(98.5\pm0.3)\%$. Although this hit efficiency is still high when taking into
 account the challenging charge to threshold ratio, it clearly shows that the
 present ATLAS read-out chip in combination with sensors of 75\,\mum\ is not
 optimal for tracking purposes. Notwithstanding the high CCE, the situation
 stays challenging after irradiation and thus a discussion of the hit
 efficiencies is not sensible and omitted here.

 The lower minimal thresholds offered by the FE-I4 read-out chip improves the
 charge to threshold ratio, and might allow to use sensors as thin as
 75\,\mum. Nonetheless, already sensors as thin as 150\,\mum\ exhibit a very
 good CCE after irradiation and are operable at comparably low bias
 voltages~\cite{TerzoIWORID}.
%
%
\section{Conclusions}
 Mechanical and electrical results obtained with SLID interconnected structures
 from an R\&D campaign towards a new pixel module were discussed. The
 investigated concept is based on several new technologies, namely n-in-p
 sensors, thin sensors, slim edges with or without active edges, and
 3D-integration incorporating {SLID} interconnections as well as {ICVs}.

 The 3D-integration is foreseen in the module concept to achieve compact
 module. The {SLID} interconnection technique by {EMFT} was qualified for use on
 pixel sensors by verifying the effectiveness of the TiW diffusion barrier and
 determining the needed vertical and horizontal alignment precision. Especially,
 it was shown that deliberate height mismatches of up to 1\,\mum\ are not
 detrimental for the connection efficiency.
 First prototype modules employing the ATLAS FE-I3 read-out chip and
 75\,\mum\ thick sensors were built. It was shown that {SLID} interconnections
 have a stability and durability similar to other interconnection technologies
 used. Furthermore, all pixel cells were interconnected for assemblies where the
 underlying {BCB} passivation layer was fully opened in correspondence to the
 {SLID} interconnections. An SF$_6$ plasma descum process will guarantee that
 interconnections the {BCB} passivation layer are opened sufficiently everywhere
 in future. 
 Also at the moment new tools and processes are installed and implemented at
 {EMFT} to further improve on the alignment precision, which will allow for even
 smaller pitches.

 For the prototype modules the {CCE} and the absolute collected charge were
 investigated systematically as functions of the received fluence and the
 applied bias voltage. 
 The results were compared to results obtained with strip sensors from the same
 production.
 It was shown that after an irradiation to a received fluence of
 $10^{16}$\,\neqcm, assemblies with a thickness of \dthin\ saturate at a {CCE}
 of 90$\%$ to 100$\%$.
 For an application within an experiment, in addition to the {CCE} also the
 absolute charge and its relation to the threshold of the read-out chip has to
 be taken into account. Although, with the low thresholds possible with the new
 FE-I4 read-out chip using a sensor thickness down to about 75\,\mum\ seems
 feasible, the absolute charge measurement indicates that a somewhat larger
 charge would be preferable to retain a good signal to threshold ratio up to the
 highest fluences expected.
 Anyhow, other factors like the lowered occupancy of thinner detectors might
 render thinner sensors still to be the better choice. Furthermore, the
 requirements on high voltage stability are relaxed for thinner sensors since
 thinner sensors exhibit a high {CCE} already at moderate bias voltages.
 In conclusion, the good properties of the sensors and modules presented here
 make them well suited for use in ATLAS when operating at the HL-LHC.
%
%
\section{Acknowledgements}
\label{sec:acknowledgment}
 This work has been partially performed in the framework of the CERN RD50
 Collaboration. The authors thank A.~Dierlamm (KIT), and V.~Cindro and I.~Mandić
 (Jo\v{z}ef-Stefan-Institut) for the sensor irradiation. Part of the irradiation
 programme was supported by the Initiative and Networking Fund of the Helmholtz
 Association, contract HA-101 ("Physics at the Terascale"). Another part of the
 irradiation and the beam test measurements leading to these results has
 received funding from the European Commission under the FP7 Research
 Infrastructures project AIDA, grant agreement no. 262025. Beam test
 measurements were conducted within the PPS beam test group comprised by:
 S.~Altenheiner, M.~Backhaus, M.~Bomben, D.~Forshaw, Ch.~Gallrapp, M.~George,
 J.~Idarraga, J.~Janssen, J.~Jentzsch, T.~Lapsien, A.~La Rosa, A.~Macchiolo,
 G.~Marchiori, R.~Nagai, C.~Nellist, I.~Rubinskiy, A.~Rummler, G.~Troska,
 Y.~Unno, P.~Weigell, J.~Weingarten.
%
%

%
%

\begin{thebibliography}{99}
\bibitem{pixelelectronics} G.~Aad et al., {\it ATLAS pixel detector electronics
  and sensors}, JINST~{\bf 3}, (2008), P07007.
\bibitem{Peric2006178} I.~Peric et al., {\it The FEI3 readout chip for the ATLAS
  pixel detector}, Nucl.~Instr.~and Meth.~{\bf A565}, (2010), 178.
\bibitem{Fritzsch2011189} T.~Fritzsch et al., {\it Cost effective flip chip
  assembly and interconnection technologies for large area pixel sensor
  applications}, Nucl.~Instr.~and Meth.~{\bf A650}, (2011), 189.
\bibitem{Lumi} L.~Rossi et al., {\it High Luminosity Large Hadron Collider: A
  Description for the European Strategy Preparatory Group}, CERN, (2012),
  CERN-ATS-2012-236.
\bibitem{IBL-TDR} M.~Capeans et al., {\it ATLAS Insertable B-Layer Technical
  Design Report}, CERN, (2010), CERN-LHCC-2010-013.
\bibitem{BenoitPhD} M.~Benoit, {\it \'Etude des d\'etecteurs planaires pixels
    durcis aux radiations pour la mise \`{a} jour du d'etecteur de vertex
    d'ATLAS}, PhD thesis, (2011), University Paris Sud - Paris XI.
\bibitem{WittigPhD} T.~Wittig, {\it Design and Quality Control of Planar ATLAS
  IBL Sensors Based on Slim Edge Studies}, PhD thesis, (2013), Technical
  University Dortmund.
\bibitem{IBL_Proto} ATLAS IBL Collaboration, {\it Prototype ATLAS IBL Modules
  using the FE-I4A Front-End Readout Chip}, JINST~{\bf 7}, (2012), P11010.
\bibitem{GarciaSciveres2010} M.~Garcia-Sciveres et al., {\it The FE-I4 pixel
  readout integrated circuit}, Nucl.~Instr.~and Meth.~{\bf A636} Supplement,
  (2011), S155.
\bibitem{LOI_II} ATLAS Collaboration, {\it Letter of Intent for the Phase-II
  Upgrade of the ATLAS Experiment}, CERN, (2012), CERN-2012-022 LHCC-I-023,
  \url{https://cds.cern.ch/record/1502664}.
\bibitem{Lacithin} L.~Andricek et al., {\it Processing of ultra-thin silicon
  sensors for future e$^{+}$e$^{-}$ linear collider experiments}, IEEE
  Trans.~Nucl.~Sci.~{\bf 51}, (2004), 1117.
\bibitem{EMFT} Fraunhofer Einrichtung für Modulare Festk\"orper-Technologie,
  \url{http://www.emft.fraunhofer.de/}.
\bibitem{TerzoIWORID} S.~Terzo et al., {\it Heavily irradiated n-in-p thin
  planar pixel sensors with and without active edges}, Proceedings of the iWoRID
  2013 Conference, JINST {\bf 9} (2014), C05023.
\bibitem{NinPpaper} C.~Gallrapp et al., {\it Performance of novel silicon n-in-p
  planar pixel sensors}, Nucl.~Instr.~and Meth.~{\bf A679}, (2012), 29.
\bibitem{WeigellPhD} P.~Weigell, {\it Investigation of properties of novel
  silicon pixel assemblies employing thin n-in-p sensors and 3D-integration},
  PhD Thesis, (2013), Technical University München, MPP-2013-5,
  CERN-THESIS-2012-229.
\bibitem{Casse2010401} G.~Casse et al., {\it Enhanced efficiency of segmented
  silicon detectors of different thicknesses after proton irradiations up to $1
  \times 10^{16}$\,n$_{\mathrm{eq}}$/cm$^2$}, Nucl.~Instr.~and Meth.~{\bf A624},
  (2010), 401.
\bibitem{Mandic2010474} I.~Mandi\'{c} et al., {\it Observation of full charge
  collection efficiency in heavily irradiated n+p strip detectors irradiated up
  to $3\times10^{15}$\,n$_{\mathrm{eq}}$/cm$^2$}, Nucl.~Instr.~and Meth.~{\bf
  A612}, (2010), 474.
\bibitem{AnnaJapan} A.~Macchiolo et al., {\it Thin n-in-p pixel sensors and the
  SLID-ICV vertical integration technology for the ATLAS upgrade at HL-LHC}, 
  Nucl.~Instr.~and Meth.~{\bf A731} (2013) 210.
\bibitem{AnnaTippSLID} A.~Macchiolo et al., {\it SLID-ICV Vertical Integration
  Technology for the ATLAS Pixel Upgrades}, Phys.~Proc.~{\bf 37}, (2012), 1009.
\bibitem{WeigellPsd} P.~Weigell et al., {\it Characterization of Thin Pixel
  Sensor Modules Interconnected with SLID Technology Irradiated to a Fluence of
  2$\cdot10^{15}$\,n$_{\mathrm{eq}}$/cm$^2$}, JINST~{\bf 6}, (2011), C12049.
\bibitem{AnnaGrindel} A.~Macchiolo et al., {\it Performance of thin pixel
  sensors irradiated up to a fluence of $10^{16}$\,n$_{\mathrm{eq}}$/cm$^2$ and
  development of a new interconnection technology for the upgrade of the ATLAS
  pixel system}, Nucl.~Instr.~and Meth.~{\bf A650}, (2011), 145.
\bibitem{BeimfordeTWEPP} M.~Beimforde et al., {\it A module concept for the
  upgrades of the ATLAS pixel system using the novel SLID-ICV vertical
  integration technology}, JINST~{\bf 5}, (2010), C12025.
\bibitem{BeimfordePhD} M.~Beimforde, {\it Development of thin sensors and a
  novel interconnection technology for the upgrade of the ATLAS pixel system},
  PhD Thesis, (2010), Technical University M\"unchen, MPP-2010-115,
  CERN-THESIS-2010-280.
\bibitem{FE-I2} L.~Blanquart et al., {\it FE-I2: a front-end readout chip
  designed in a commercial 0.25-\mum\ process for the ATLAS pixel detector at
  LHC}, IEEE Trans.~Nucl.~Sci.~{\bf 51}, (2004), 1358.
\bibitem{stockmanns} T.~Stockmanns, {\it Multi-Chip-Modul-Entwicklung f\"ur den
  ATLAS-Pixeldetektor}, PhD Thesis, (2004), Bonn University.
\bibitem{Weigell2011} P.~Weigell et al., {\it Characterization and performance
  of silicon n-in-p pixel detectors for the ATLAS upgrades}, Nucl.~Instr.~and
  Meth.~{\bf A658}, (2011), 36.
\bibitem{Bernstein1966} L.~Bernstein et al., {\it Applications of Solid-Liquid
  Inderdiffusion (SLID) Bonding in integrated-Circuit Fabrication},
  Trans.~Met.~Soc.~AIME~{\bf 236m}, (1966), 405.
\bibitem{Bernstein1965} L.~Bernstein, {\it Semiconductor brazing by the
  solid-liquid-inter-diffusion (SLID) process}, in: ECS Meeting, San Francisco,
  (1965), 319.
\bibitem{Klumpp3DBuch} A.~Klumpp, {\it Bonding with Intermetallic Compounds} in
  P.~Garrou et al., {\it Handbook of 3D Integration: Technology and Applications
    of 3D Integrated Circuits}, Wiley-VCH, Weinheim(Germany), (2008), 261
\bibitem{Klumpp20mum} H.~H\"ubner et al., {\it Face-to-Face Chip Integration
  with Full Metal Interface}, in B.~Melnick et al., {\it Advanced Metallization
  Conference Proceedings}~{\bf XVIII}, (2002), 53.
\bibitem{3DFermi} G.~Deptuch et al., {\it 3D Technologies for Large Area
  Trackers}, Whitepaper Submitted to Snowmass 2013,
  \url{http://arxiv.org/pdf/1307.4301.pdf}
\bibitem{LETI-copper} S.~Joblot et al., {\it Copper pillar interconnect
  capability for mmwave applications in 3D integration technologies},
  Microelectronic Engineering~{\bf 107}, (2013), 72.
 \bibitem{LETI} Leti, \url{http://http://www-leti.cea.fr}.
\bibitem{keithley} Keithley Instruments Inc, \url{http://www.keithley.com/}.
\bibitem{Istratov} A.A.~Istratov and E.R.~Weber, {\it Physics of Copper in
  Silicon}, Journal of the Electrochemical Society~{\bf149~(1)}, (2002), G21.
\bibitem{pixeltdr} ATLAS Collaboration, {\it Pixel Detector Technical Design
  Report}, CERN, (1998), CERN-LHCC-98-013.
\bibitem{EMFTpriv} A.~Klumpp, EMFT, private communication.
\bibitem{IZM} Fraunhofer Institut f\"ur Zuverl\"assigkeit und Mikrointegration,
  \url{http://www.izm.fraunhofer.de/}.
\bibitem{BumpGoPatent} T.~Go, {\it Bonding of aligned conductive bumps on
  adjacent surfaces}, US Patent 4912545, (1987).
\bibitem{Broennimann2006303} Ch.~Broennimann et al., {\it Development of an
  Indium bump bond process for silicon pixel detectors at PSI}, Nucl.~Instr.~and
  Meth.~{\bf A565}, (2006), 303.
\bibitem{Eldring} J.~Eldring et al., {\it Flip Chip Attach of Silicon and GaAs
  Fine Pitch Devices as well as Inner Lead TAB Attach Using Ball-bump
  Technology}, Microelectron Int.~{\bf 11}, (1994), 20.
\bibitem{Broennimann:gf0003} Ch.~Broennimann et al., {\it The PILATUS 1M
  detector}, J.~Synch.~Rad.~{\bf 13}, (2006), 120.
\bibitem{Cheah} L.~Cheah et al., {\it Gold to gold thermosonic flip-chip
  bonding}, SPIE Proc.~Series~{\bf 4428}, (2001), 165.
\bibitem{AldiPhD} A.~Dierlamm, {\it Untersuchungen zur Strahlenhärte von
  Siliziumsensoren}, PhD Thesis, (2003), Karlsruhe University, IEKP-KA/2003-23.
\bibitem{Furgeri} A.~Furgeri, {\it Qualit\"atskontrolle und Bestrahlungsstudien
  an CMS Siliziumstreifensensoren}, PhD Thesis, (2006), Karlsruhe University,
  IEKP-KA/2005-1.
\bibitem{Snoj2012483} L.~Snoj et al., {\it Computational analysis of irradiation
  facilities at the JSI TRIGA reactor}, Appl.~Rad.~Iso.~{\bf 70}, (2012), 483.
\bibitem{MollPhD} M.~Moll, {\it Radiation Damage in Silicon Particle Detectors},
  PhD Thesis, (1999), Hamburg University.
\bibitem{USBPix} USB based readout system for ATLAS FE-I3 and FE-I4,
  \url{http://icwiki.physik.uni-bonn.de/twiki/bin/view/Systems/UsbPix}.
\bibitem{BichselPDB} H.~Bichsel et al., {\it Passage of Particles through
  matter}, J.~Phys.~G~{\bf 637}, (2010), 285.
\bibitem{Bichsel1988} H.~Bichsel, {\it Straggling in thin silicon detectors},
  Rev.~Mod.~Phys.~{\bf 60}, (1988), 663.
\bibitem{MaltePSD9} M.~Backhaus, {\it Characterization of new hybrid pixel
  module concepts for the ATLAS Insertable B-Layer upgrade}, JINST~{\bf 7},
  (2012), C01050.
\bibitem{Am241_pub} R.~Gehrke et al., {\it Radioactinide additions to the
  electronic Gamma-ray Spectrum Catalogue}, J.~Rad.~Nucl.~Chem.~{\bf 248},
  (2001), 417.
\bibitem{Am241_spectrum} Ray Spectrometry Center, {\it Gamma-Ray spectrum
  catalogue Idaho National Engineering \& Environmental Laboratory}, (2001).
\bibitem{BONN-IR-2008-04} J.~Gro{\ss}e-Knetter, {\it Vertex Measurement at a
  Hadron Collider, The ATLAS Pixel Detector}, Habilitation thesis, Bonn
  University, (2008), BONN-IR-2008-04.
\bibitem{EUDET} I.~Rubinskiy, {\it An EUDET/AIDA Pixel Beam Telescope for
  Detector Development}, Phys.~Proc.~{\bf 37}, (2012), 923.
\bibitem{Paterno} M.~Paterno, {\it Calculating Efficiencies and Their
  Uncertainties}, FERMILAB, (2004), FERMILAB-TM-2286-CD.
\end{thebibliography}
\end{document}